\documentclass[10pt,aps,prd,twocolumn,superscriptaddress,amsmath, nofootinbib]{revtex4-1}

\usepackage{graphicx}
\usepackage{dcolumn}
\usepackage{bm}

\usepackage{capt-of}
\usepackage{placeins}
\usepackage{aas_macros}
\usepackage{lipsum}
\usepackage{tikz}
\usepackage{layouts}
\usepackage{colortbl}
\usepackage[hidelinks]{hyperref}
\usepackage[capitalize]{cleveref}
\Crefname{equation}{Equation}{Equations}
\crefname{equation}{Eq.}{Eqs.}
\Crefname{figure}{Figure}{Figures}
\crefname{figure}{Fig.}{Figs.}
\Crefname{table}{Table}{Tables}
\crefname{table}{Tab.}{Tabs.}
\Crefname{section}{Section}{Sections}
\crefname{section}{Sec.}{Secs.}

\usepackage{amsthm}
\theoremstyle{definition}

\crefname{proposal}{Proposition}{Proposition}

\usepackage{siunitx}
\DeclareSIUnit\parsec {pc}
\DeclareSIUnit \h {\ensuremath{\mathit{h}}}
\DeclareSIUnit \solarmass {\ensuremath{\mathit{M_{\odot}}}} 
\DeclareSIUnit\arcsec {as}
\DeclareSIUnit\year {yr}

\newcommand{\Mlg}{$M_{\rm MW + M31}$}
\def\P{\mathcal{P}}
\def\L{\mathcal{L}}
\def\Z{\mathcal{Z}}
\newcommand{\gaia}{\textit{Gaia}}

\begin{document}

\title{The sum of the masses of the Milky Way and M31: a likelihood-free inference approach}

\author{Pablo Lemos}
\email[]{pablo.lemos.18@ucl.ac.uk}
\affiliation{Department of Physics and Astronomy, University College London, Gower Street, London, WC1E 6BT, UK}

\author{Niall Jeffrey}
\affiliation{Laboratoire de Physique de l'Ecole Normale Sup\'erieure, ENS, Universit\'e PSL, CNRS, Sorbonne Universit\'e, Universit\'e de Paris, Paris, France }
\affiliation{Department of Physics and Astronomy, University College London, Gower Street, London, WC1E 6BT, UK}

\author{Lorne Whiteway}
\affiliation{Department of Physics and Astronomy, University College London, Gower Street, London, WC1E 6BT, UK}

\author{Ofer Lahav}
\affiliation{Department of Physics and Astronomy, University College London, Gower Street, London, WC1E 6BT, UK}

\author{Noam I Libeskind}
\affiliation{Leibniz-Institut für Astrophysik Potsdam (AIP), An der Sternwarte 16, 14482 Potsdam, Germany}
\affiliation{University of Lyon, UCB Lyon-1/CNRS/IN2P3, IPN Lyon, France}

\author{Yehuda Hoffman}
\affiliation{Racah Institute of Physics, Hebrew University, Jerusalem, 91904 Israel}

\begin{abstract}
We use Density Estimation Likelihood-Free Inference,  
$\Lambda$ Cold Dark Matter
simulations of $\sim 2M$ galaxy pairs, and data from \gaia{} and the Hubble Space Telescope
to infer 
the sum of the masses of the Milky Way and Andromeda (M31) galaxies, the two main components of the Local Group. 
This method overcomes most of the approximations of the traditional timing
argument, makes the writing of a theoretical likelihood unnecessary, and allows the non-linear modelling of
observational errors that take into account correlations in the data and non-Gaussian distributions. We obtain an $M_{200}$ mass estimate
$M_{\rm MW+M31} =$ \SI[parse-numbers=false]{4.6^{+2.3}_{-1.8} \times 10^{12}}{\solarmass}
($68 \%$ C.L.), in agreement with previous estimates 
both for the sum of the two masses and for the individual masses.
This result is not only
one of 
the most reliable estimates of the sum of the two masses
to date, but is also an illustration of
likelihood-free inference
in a problem with only
one parameter and only three data points.
\end{abstract}


\maketitle


\section{Introduction}
\label{sec:intro}

Likelihood-free inference (LFI) has emerged as a very promising technique for inferring parameters from data,
particularly in cosmology. It provides parameter posterior probability estimation without requiring the calculation of an analytic 
likelihood (i.e. the probability of the data being observed given the parameters).
LFI uses forward simulations in place of an analytic likelihood function. Writing a likelihood for cosmological observables can be
extremely complex, often requiring the solution of Boltzmann equations, as well as approximations for highly nonlinear 
processes such as structure formation and baryonic feedback. While simulations have their own 
limitations and are computationally expensive, the quality and efficiency of cosmological simulations are 
constantly increasing, and they are likely to soon far surpass the accuracy or robustness of any likelihood function. 

This is a rapidly growing topic in cosmology, due to the emergence of novel methods for likelihood-free inference \citep[see e.g.][]{Leclercq2018, Alsing2019}, with applications to data sets such as the Joint Light Curve (JLA) and Pantheon supernova datasets \citep{Betoule2014, 2020arXiv200510628W}, and the Dark Energy Survey Science Verification data~\citep{Jeffrey2020}, amongst others \citep{Brehmer:2019jyt, Ramanah:2020ylz, 2020JCAP...09..048T}. There are, therefore, many applications for which LFI could improve the robustness of parameter inference using cosmological data. In this work we perform a LFI-based parameter estimation 
of the sum of masses of the Milky Way and M31. 
The likelihood function for this problem requires significant simplifications, but forward simulations can be
obtained easily.




The Milky Way and Andromeda are the main components of the Local Group, which includes tens of smaller galaxies.
We define \Mlg{} as the sum of the MW and M31 masses.
Estimating \Mlg{} remains an elusive and complex problem in astrophysics. As the mass of each of the
Milky Way and M31 is known only to within a factor of 2, it is important to constrain the sum of their masses. The
traditional approach is to use the so-called timing argument (TA) \citep{Kahn1959}. The timing argument estimates
\Mlg{} using Newtonian Dynamics
integrated from the Big Bang. This integration is an extremely simplified version of a very complex problem. Therefore, 
alternative methods that do not rely on the same approximations become extremely useful. 

In this work, we use the {\it MultiDark Planck} (MDPL) simulation\footnote{\url{www.cosmosim.org}} 
\cite{Prada2012, 2013AN....334..691R}, combined with data from the Hubble Space Telescope \citep[HST, ][]{HST} and \gaia{} \citep{Gaia}, to estimate \Mlg{}.
A similar data set was previously used in \citep{McLeod2017} to obtain
a point estimate of \Mlg{} using Artificial Neural Networks (ANN) in conjunction with the TA. In contrast, our work uses Density Estimation 
Likelihood-Free Inference \citep[DELFI][]{Bonassi2011, Fan2012, Papamakarios2016, Alsing2019}, 
using the {\tt pyDELFI} package\footnote{
\url{https://github.com/justinalsing/pydelfi}
}, combined with 
more recent data. While the result is important on its own, this paper also illustrates the fundamental methodology
of DELFI in a problem that is statistically simple but physically complex.

The structure of the paper is as follows: \cref{sec:previous} reviews and describes previous estimates of \Mlg{}. \cref{sec:lfi} describes the basics of LFI, and the particular techniques used in this work. 
\cref{sec:simulations} and \cref{sec:data} describe the simulations and data, respectively, used in this work. 
\cref{sec:results} shows our results, and conclusions are presented in \cref{sec:conclusions}.

\section{Previous Estimates}
\label{sec:previous}

\begin{table*}
	\centering
	\caption{Estimates of \Mlg{} from previous work. The third column shows the data used, with
 $r$ in \si{\mega\parsec} and $v_r, v_t$ in \si{\kilo\metre\per\s}. The fourth column shows \Mlg{} in units of \si{10^{12} \solarmass}. 
 Note
that Gaussian approximations have been used to convert the reported confidence levels to $68\%$ confidence levels in some cases. }
	\label{tab:example_table}
	\renewcommand{\arraystretch}{1.3} 
	\begin{tabular}{lccr} 
		\hline
		Reference & method & assumed  ($r, v_r, v_t$  )& \Mlg{} \\
		\hline
		Li \& White (2008) \citep{Li:2007eg} & TA calibrated on Sims & (0.784, 130, 0) & $5.27^{+2.48}_{-0.91}$ \\
		Gonzalez \& Kravtsov (2014) \citep{Gonzalez:2013pqa} & Sims & (0.783, 109.3, 0) & $4.2^{+2.1}_{-1.2}$ \\
		McLeod et al. (2017) \citep{McLeod2017} & TA+$\Lambda$ & ($0.77\pm 0.04, 109.4 \pm 4.4, 17\pm 17$)    &$4.7 ^{+0.7+2.9}_{-0.6-1.8}$ \\
		McLeod et al. (2017) \citep{McLeod2017} & Sims + ANN +Shear &    ($0.77\pm 0.04, 109.4 \pm 4.4, 17\pm 17$)    &$4.9 ^{+0.8+1.7}_{-0.8-1.3}$ \\
		Phelps et al. (2013) \citep{Phelps:2013rra} & Least Action  & (0.79, 119, 0) & $6.0 \pm 0.5$ \\

		\hline
	\end{tabular}
\end{table*}



A first approach to estimating \Mlg{} from dynamics, known as TA, is based on the simple idea that MW
and M31 are point masses approaching each other on a radial orbit that obeys
\begin{equation}\label{LGTA}
\ddot{r} = -\frac{GM_{\rm MW+M31}}{r^2} +  \frac{1}{3}\Lambda  \, r,
\end{equation}
where \Mlg{} is the sum of the masses of the two galaxies \citep{Kahn1959, Lynden-Bell1981},
and where the $\Lambda$ term, which represents a form of dark energy, was added in later studies \citep{Binney1987, Partridge:2013dsa}. 
It was also extended for modified gravity models
\citep{2019arXiv190310849M}.
Since we know the present-day distance $r$
between MW and M31 and their relative radial velocity $v_r$,
and if we assume the age of the universe and $\Lambda$, we can infer the mass \Mlg.
The analysis can be
extended to cover the case of non-zero tangential $v_t$ velocity \citep[e.g. ][]{McLeod2017,2019arXiv190403153B}.
The pros and cons of the timing argument are well known.
It is a simple model which assumes only two point mass bodies; this  ignores, for example, the tidal forces due to neighbouring galaxy haloes in the Local Group and the extended cosmic web around it.
While the timing argument model does not capture the complexity of the cosmic structure and resulting cosmic variance, it gives a somewhat surprisingly good estimate for \Mlg. As shown below, it can also serve to test the sensitivity 
of the results to parameters such as the cosmological constant $\Lambda$ 
and the Hubble constant $H_0$, in case simulations are not available for different values of these parameters.

A second approach is to consider the dynamics of all the galaxies in the Local Group using  
the Least Action Principle \citep{Peebles1994}, as, for example, was implemented in \cite{Phelps:2013rra}.
In this approach all 
members of the Local Group appear in the model,
but as a result the derived masses are correlated, and the error bars should be interpreted accordingly.
A third approach is to use N-body simulations of the local universe, 
assuming a cosmological model such as {$\mathrm{\Lambda CDM}$} \citep{Li:2007eg, Gonzalez:2013pqa, McLeod2017}. 
In this paper we apply this third method, but using the DELFI method (this provides significant 
improvements over other methods, as discussed in the following section).

Representative results for \Mlg{} from previous works are given in \cref{tab:example_table}.
Throughout the paper we quote $68 \%$ credible intervals.


\section{Likelihood-Free Inference}
\label{sec:lfi}


In Bayesian statistics we often face the following problem: given observed data $D_{\rm obs}$, and a 
theoretical model $I$ with a set of parameters $\theta$, calculate the probability of the 
parameters given the data. In other words, we want to calculate the {\it posterior distribution} 
$\P \equiv p(\theta | D_{\rm obs}, I)$; here $p$ is a probability (for a model with discrete parameters) 
or a probability density (for continuous parameters). 
We do so using Bayes' theorem:

\begin{equation}
\label{eq:bayes}
p(\theta | D_{\rm obs}, I) = {p(D_{\rm obs} | \theta, I) p(\theta | I) \over  p(D_{\rm obs} | I)} 
\Leftrightarrow \P = {\L \times \Pi \over \Z}
\end{equation}
where $\L$ is called the likelihood, $\Pi$ the prior, and $\Z$ the Bayesian Evidence. The Bayesian Evidence acts as 
an overall normalization in parameter estimation, and can therefore be ignored for this task. Thus, given a choice of prior 
distribution and a likelihood function, we can estimate the posterior distribution.
However, obtaining a 
likelihood function is not always easy. The likelihood function provides a probability of measuring the
data as a function of the parameter values, and often requires approximations
both in the statistics and in the theoretical modelling. 

Likelihood-Free Inference is an alternative method for calculating the posterior distribution;
in this method we do not formally write down a likelihood function.
Instead, we use {\it forward simulations} of the model to generate samples 
of the data and parameters. In the simplest version of LFI, we select only the forward simulations that 
are the most similar to the observed data, rejecting the rest. This method is known as Approximate Bayesian 
Computation \citep[ABC, ][]{Rubin1984}; it relies on choices of a distance metric (to measure similarity between 
simulated and observed data) and of a maximum distance parameter $\epsilon$ (used to accept or rejected simulations).

In this work, we will use a version of LFI called Density Estimation Likelihood-Free Inference (DELFI). In this 
approach, we use all existing forward simulations to learn a conditional density distribution of the 
data\footnote{Note that we use the letter $d$ to refer to the data space used in DELFI to learn conditional
distributions, and we use $D_{\rm obs}$ to refer to the observed data.
}
$d$ given the parameters $\theta$, using a density estimation algorithm. Examples of density estimation algorithms 
are Kernel Density Estimation \citep[KDE, ][]{rosenblatt1956, parzen1962, simonoff1996smoothing}, Mixture Models, 
Mixture Density Networks \citep{Bishop94mixturedensity, Bishop2006}, and Masked Autoregressive Flows 
\citep{Papamakarios2017}. 
We use the package {\tt pyDELFI}, and estimate the likelihood function from the
forward simulations using Gaussian Mixture Density Networks (GMDN) and Masked Autoregressive Flows (MAF).
In this sense, the name Likelihood-Free Inference is perhaps misleading: the inference is not ‘likelihood-free’, we simply avoid writing a likelihood and instead model it using forward simulations. A more accurate name for the method could therefore be ‘explicit-likelihood-free’.
A more extended discussion of our choice of density estimation algorithms and conditional distribution is
presented in \cref{sec:density}.


DELFI has several advantages over the simpler ABC approach 
to LFI: it does not rely on a choice of a distance parameter $\epsilon$ (although admittedly the choice of basis in parameter space can change the implicit distance metric of the density estimator) and it uses all 
available forward simulations to build the conditional distribution, making it far more efficient. 

While relatively new, likelihood-free inference has already been applied to several problems in astrophysics
\citep[e.g. ][]{Cameron2012, Weyant2013, Akeret:2015uha, Hahn:2016zwc, Peel:2016jub, Kacprzak:2017nxl, Jeffrey2020}. 
However, most applications involving LFI suffer from the curse of dimensionality: there can be hundreds,
thousands, or even millions of observables (such as $\sim 2000$ multipoles in Cosmic Microwave Background surveys, 
or $~500$ redshift and angular bins in cosmic shear analyses), and it is impossible to perform density estimation. Some form of data compression is therefore usually needed
\citep{Alsing2018, Alsing2018b, heavens2020extreme}. 
Similarly, due to the high dimensionality and complexity of these parameter spaces, efficient methods to 
generate the simulations (so as to minimize the number needed) have been developed 
\citep{papamakarios2019sequential, lueckmann2019likelihood,Alsing2019}.  
However, our \Mlg{} problem has only three data points and one parameter of interest, making it an extremely
simple application of the method from the statistical point of view; it illustrates all necessary
techniques, and does not require data compression.

To summarize, the steps that we will follow are: 

\begin{itemize}
\item Generate a large number of simulations of systems similar to the one of interest. The simulations used 
in this work are described in \cref{sec:simulations}.
\item Use a density estimator, in our case GMDN and MAF as part of the \texttt{pyDELFI} package, to obtain the sampling distribution 
for any data realization
$p(d | \theta, I)$.
\item Evaluate this distribution at the observed data (which will be described in \cref{sec:data}) to 
obtain the likelihood function for our observed data realization: $p(d = D_{\rm obs} | \theta, I)$.
\item From a prior distribution and the likelihood, and using Bayes' theorem (\cref{eq:bayes}), get a posterior 
distribution $p(\theta | D_{\rm obs}, I)$.
\end{itemize}


\section{Simulations} \label{sec:simulations}
We use the publicly available MDPL
simulation. 
This is an $N$-body
simulation of a periodic box with side length $L_{\rm box}=$ \SI{1}{\giga\parsec\per\h},
populated with $2048^3$ dark
matter particles. Such a simulation achieves a mass resolution of \SI{8.7E9}{\solarmass\per\h} and a
Plummer equivalent gravitational softening of \SI{7}{\kilo\parsec}. The simulation was run with the ART
code and assumed a $\Lambda$CDM power spectrum of fluctuations
with $\Omega_{\Lambda}=0.73$, $\Omega_{\rm b}=0.047$,  $\Omega_{\rm m}=0.27$,
$\sigma _{\rm 8}= 8.2$ and $H_{0}= \SI{100}{\h\kilo\metre\per\s\per\mega\parsec}$, with $h=0.7$. 

Haloes are identified with the AHF halo finder \citep{Knebe2011}. AHF identifies local
over-densities in the density field as possible halo centres. The local potential
minimum is computed for each density peak and the particles that are gravitationally
bound are identified. Haloes with more than 20 particles are recorded in the halo
catalogue. The AHF catalogue includes some 11,960,882 dark matter haloes.

The halo catalogue is then searched for pairs of galaxies as follows.
We first identify those haloes in the mass\footnote{
Henceforth, all masses are defined as $M_{200}$, the total mass enclosed in the largest
sphere surrounding them with an enclosed mean density over
200 times the critical value \citep{Li:2007eg}.
} range $5\times10^{10} < M/M_{\odot}h^{-1} < 5\times10^{13}$.
For each such halo, we find all haloes (irrespective of mass) within $\sim$\SI{4}{\mega\parsec} and sort these by distance.
If the closest of these is within the mass range $[5\times10^{10},5\times10^{13}]$ and is separated by more
than \SI{500}{\kilo\parsec} but less than \SI{1500}{\kilo\parsec}, the two haloes are considered a potential pair. The partner is then
examined to ensure that the pair is isolated (namely that there is no other halo with a mass
$>5\times10^{10}$ and closer to either pair member than the pair separation). If this is the case then
the pair is kept for the analysis described here. In this way, 1,094,839 pairs are found (as opposed
to the $30,190$ pairs used in \citep{McLeod2017}). 

We believe the criteria used to select the halo pairs in this work are not very restrictive and any cuts lie well outside the realistic limits for the actual MW-M31 system. Though these selections are unrestrictive, we note that including more pairs does improve the density estimation, as DELFI can use all the available simulated data, and not only those that are close to the observation. 


\section{Data}
\label{sec:data}

\begin{figure}
  \centering
    \includegraphics[width=0.99\columnwidth]{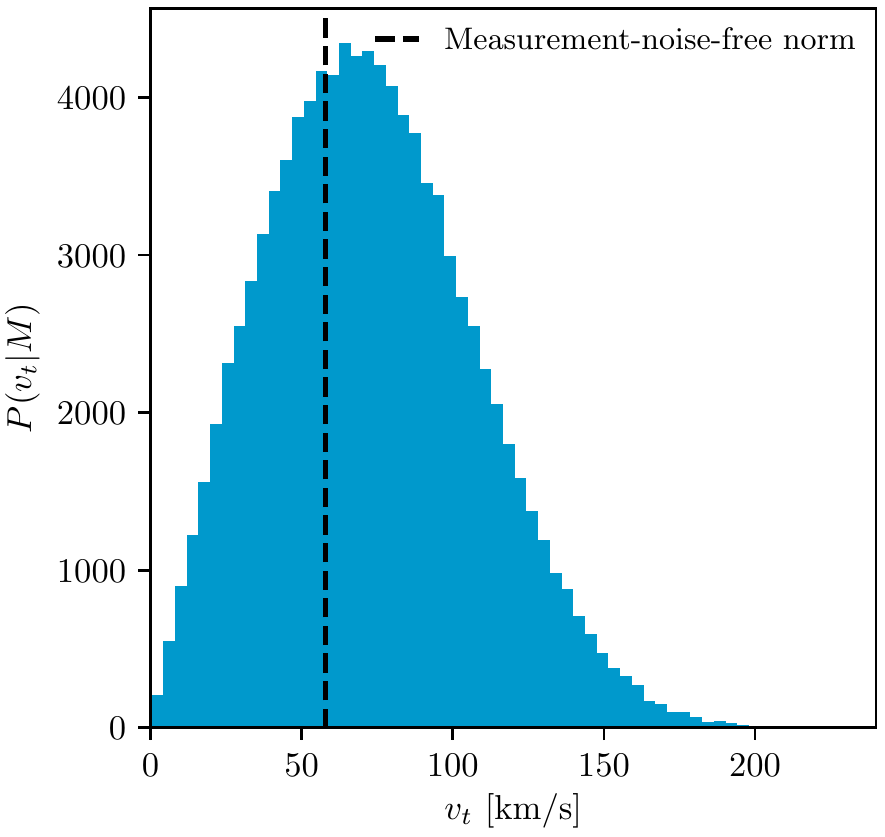}
  \caption{An illustration of our non-Gaussian error modelling for the tangential velocity. The plot was obtained taking 
  the components of the tangential velocity for a 
  randomly chosen simulation (with fixed $M$), scattering a large number of times by the errors of \cref{eq:mu}, and calculating the norm
 for each sample.
 The black dashed line shows the norm of the components for the tangential velocity of the simulation without measurement error.
\label{fig:vt}}
\end{figure}




We use three observations to constrain \Mlg{}: the distance to M31, and the radial and tangential components of its
velocity. While the TA requires other observables, such as the age of the Universe $t_0$ and the cosmological constant
$\Lambda$, in our approach these are already included in the simulations at fixed values. 
We discuss below how uncertainties in cosmological parameters affect the estimated mass. 

For the distance to M31, we adopt the commonly used value $r =$ \SI{770 \pm 40}{\kilo\parsec}
\citep{Holland:1998br, Joshi:2002uf, Ribas:2005uw, McConnachie:2005td}. For the velocity, we follow the results of 
\citep[][ henceforth VdM19]{vdm2019}. The radial velocity in the galactocentric rest frame is
$v_r =\SI{-109.4 \pm 4.4}{\kilo\metre\per\s}$ \citep{vdm2012} from HST 
observations. The tangential velocity is slightly more cumbersome. VdM19 report the following value for components
of the tangential velocity of M31 from a combination of \gaia{} DR2 and HST:

\begin{equation}
\label{eq:mu}
\mu = \left( 10 \pm 11, -16 \pm 11 \right) \si{\micro\arcsec\per\year},
\end{equation}
already in the galactocentric frame, i.e. after correcting for the solar reflex motion. VdM19 uses the distance to M31 to convert 
this to \si{\kilo\metre\per\s}. They then use a method described in \citep{vanderMarel:2007yw}
to correct for the fact that taking the
norm of the two components leads to a `bias' in the reported tangential velocity\footnote{
We will discuss this supposed bias in a future publication.}.
However, in this work, we take a different approach: for each simulation, we scatter the value
of each component of the tangential velocity for the 
simulation  according the observational error of each components shown in 
\cref{eq:mu}. By doing this, we are putting the observational errors in the simulated measurements, instead attempting to
`debias' the tangential velocity summary statistic. 
We convert to \si{\kilo\metre\per\s} using the value of the distance for that sample, to account for the covariance
between $r$ and $v_t$. 
We take as the observed value the norm of the two observed components, 
$v_t =$ \SI{72}{\kilo\metre\per\s}. While this differs from the value reported in VdM19,
this should not be a problem, as long as the way in which we calculate the tangential 
velocity in simulations and observations is consistent, and we use a summary statistic
that extracts all the available information (which the norm of the components does). Furthermore, \cref{sec:vt}
shows that our results do not significantly change if we repeat the analysis using a purely radial motion ($v_t = 0$).

This approach takes into account both the non-Gaussian errors in $v_t$ and the correlation of $v_t$ with the
errors in the distance measurement (these have not been accounted for in previous estimates of \Mlg{}).





\section{Results}
\label{sec:results}

\subsection{Overview}

\begin{figure*}
  \centering
    \includegraphics[width=0.99\columnwidth]{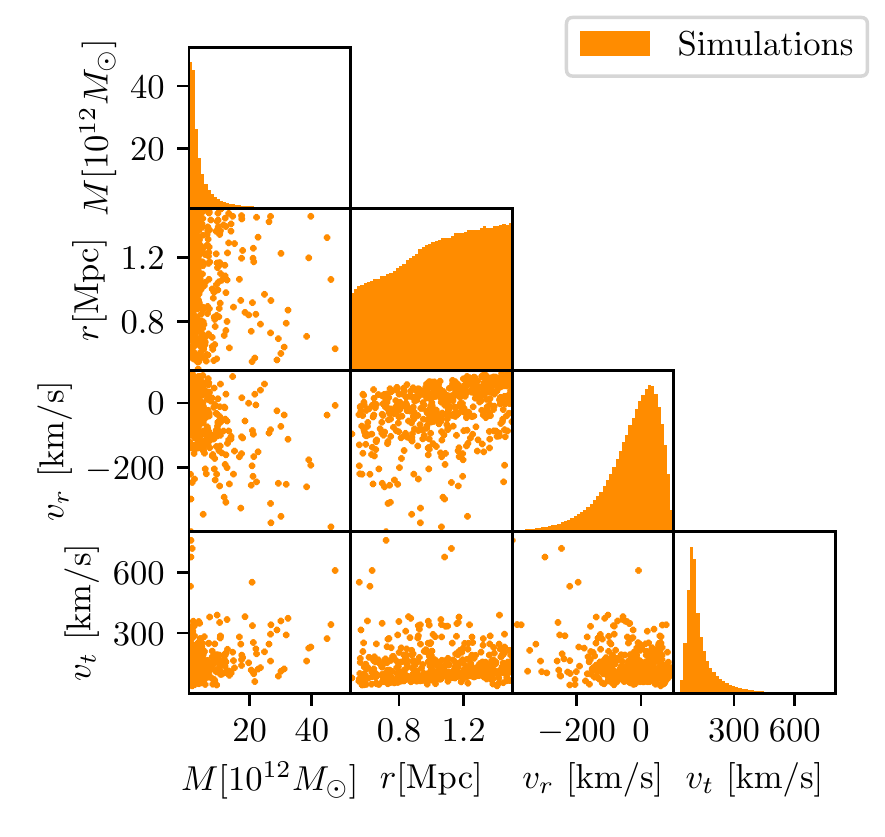}
    \includegraphics[width=0.99\columnwidth]{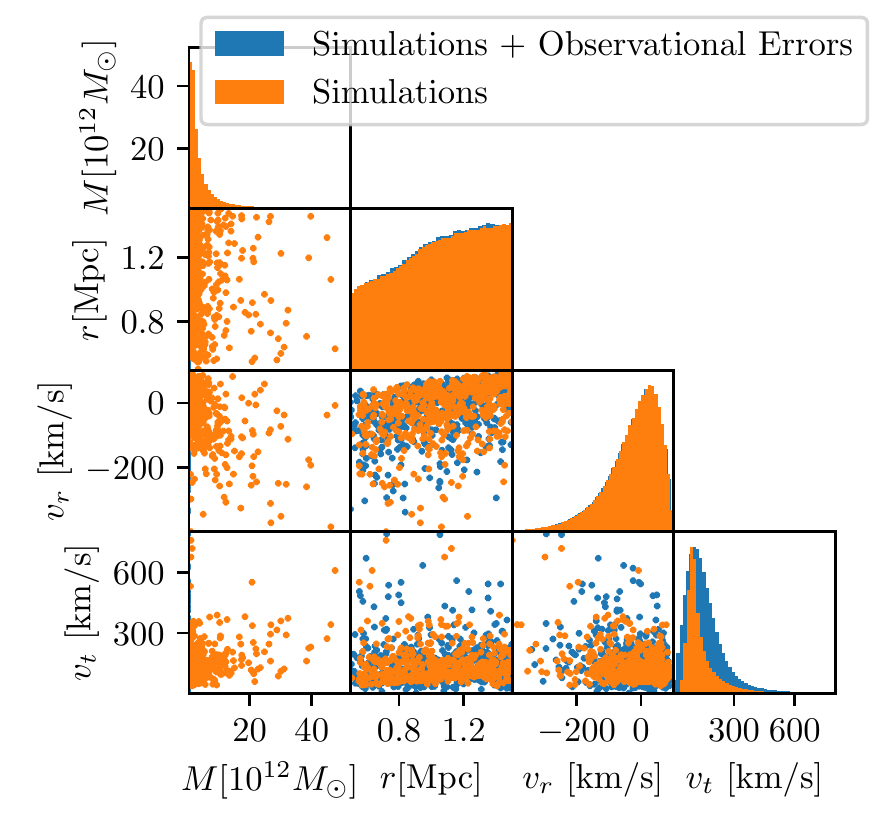}
  \caption{Illustration of DELFI for the estimation of \Mlg{}. The left panel is a scatter plot 
of the simulations described in \cref{sec:simulations}. The right panel adds the observational errors 
by scattering the simulations. 
\label{fig:delfi}
}
\end{figure*} 

\begin{figure}
  \centering
    \includegraphics[width=0.99\columnwidth]{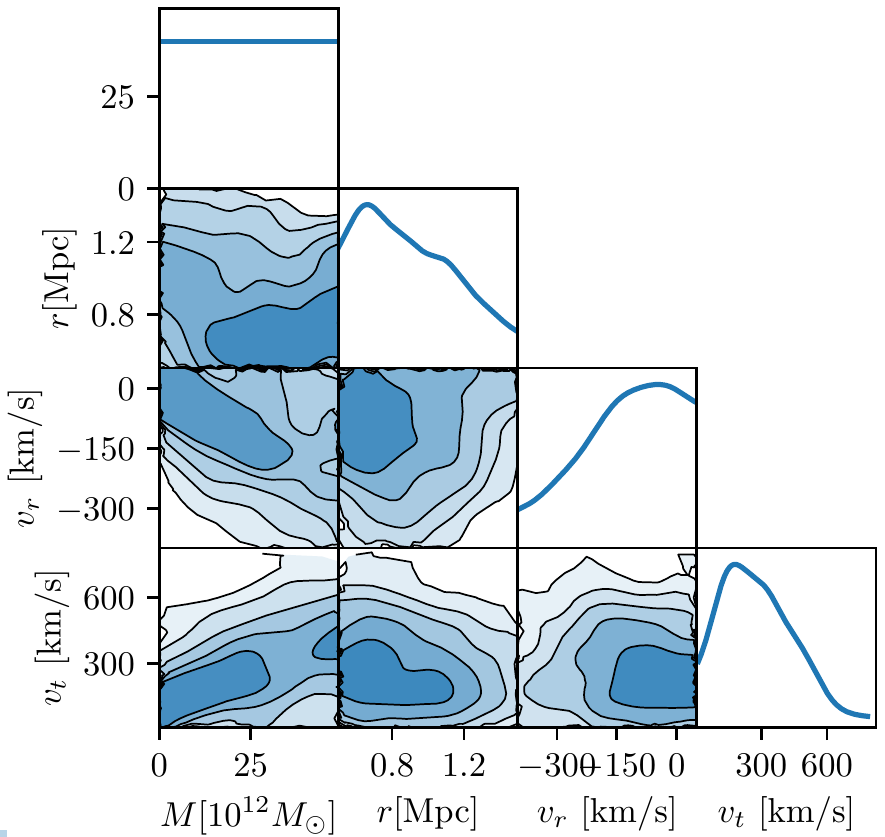}
  \caption{Conditional distribution $p(d | \theta, I)$ density estimate from the points shown in the right panel of \cref{fig:delfi}. The 2D subplots in the left column show the probabilities of the data $d = \left\{ r, v_r, v_t \right\}$ conditional on the parameter $\theta = M$ (which has been given a uniform distribution), while the remaining 2D subplots have been marginalised over this uniform distribution. By evaluating this function at the observed data points $d = \left\{ r, v_r, v_t \right\}$ we obtain the likelihood function. The points were sampled using the Nested Sampling \citep{skilling2006nested} code {\tt PolyChord} \citep{Handley:2015a, Handley:2015b}. Note that the one dimensional distribution on $M$ is flat by construction, as this is the parameter, and the distributions on $r$ and
$v_r$ appear to be `cut' because of the selection criteria described in \cref{sec:simulations}.
\label{fig:delfi2}
}
\end{figure}

Having discussed the method, the simulations, and the data, we have everything we need to perform LFI using DELFI. 
We have three data points $d = \left\{ r, v_r, v_t \right\}$ and one parameter $\theta = M$. 
The process is illustrated in \cref{fig:delfi} and \cref{fig:delfi2}, 
and consists of the following:

\begin{enumerate}
\item Left panel of \cref{fig:delfi}: We generate a large number of forward simulations, as discussed in \cref{sec:simulations}. 
Increasing the 
number of simulations will increase the accuracy of the density estimation, and of the resulting posterior. 
Reference \citep{Alsing2019} demonstrates an \textit{active learning} scheme with \texttt{pyDELFI}, providing criteria to run new simulations based on discrepancies between the density estimates in the neural density estimator ensemble. However, due to our initial large number of simulations, we had no need to run such extra simulations on-the-fly. 
\item Right panel of \cref{fig:delfi}: The observational errors are introduced as scatter in the forward simulations. More specifically, 
we displace the simulations by a number sampled from the error model presented in \cref{sec:data}.
Note how in \cref{fig:delfi} this step does not affect the mass. This is because the 
mass in this problem is part of the parameters $\theta$, not the data, as it is our goal to obtain a posterior 
distribution for the mass. 
\item We use density estimation to get the conditional density distribution $p(d | \theta, I)$, as shown in \cref{fig:delfi2}. 
While it might seem counter-intuitive to learn the likelihood instead of directly learning the 
posterior, this allows us to then sample from a chosen prior, instead of being limited to the prior that is 
implicit in the simulations. This is discussed in more detail in \cref{sec:density}.

There are several algorithms that can be used to get a conditional probability distribution from samples. 
In this work we use GMDN and MAF\footnote{
Note that while this work uses GMDN and MAF for density estimation, \cref{fig:delfi2} uses KDE instead. 
This is because the plots were generated using the code {\tt anesthetic} \citep{anesthetic}, which uses KDE to plot 
smooth probability distributions. KDE is appropriate in this case, as {\tt anesthetic} only plots the one- and 
two-dimensional posterior distributions, whereas in this work we are trying to learn the full 4D distribution. 
{\tt anesthetic} uses the {\tt fastKDE} implementation \citep{fastkde1, fastkde2}.
} (as part of the \texttt{pyDELFI} package). 


\item Finally, we evaluate this conditional density distribution at the observed data 
\begin{multline}
D_{\rm obs} = \{ r = \SI{0.77}{\mega\parsec}, \\ v_r = \SI{-109.3}{\kilo\metre\per\s}, v_t = \SI{72}{\kilo\metre\per\s} \},
\end{multline}
as discussed in \cref{sec:data}. This way, we get the likelihood function:

\begin{equation}
\L \equiv p(d = D_{\rm obs} | \theta, I).
\end{equation}
\end{enumerate}

Through this process we obtain a likelihood function, without ever having to write a theory or use a Gaussian 
approximation. 
While this process
is limited by the number and accuracy of the available simulations, it has a big advantage over likelihood-based 
problems that use simplifying approximations to make the likelihood more tractable, or easier to compute. The 
calculation of \Mlg{} is a good example: the likelihood-based approach relies on the TA and data modelling approximations, which we know
oversimplifies the problem. Instead, using DELFI, we can account for the complex nonlinear evolution of the system 
through our N-body simulation. 

\subsection{Density estimation validation}

For the density estimation, with \texttt{pyDELFI} we use a combination of two GMDNs (with four and five Gaussian components) and two MAFs 
(with three and four components). The GMDNs have two layers with fifty components each, while the MAFs have thirty 
components on each of the two hidden layers.
For a more robust density estimation we stack the results weighted by each density estimation's relative 
likelihood, as described in \citep{Alsing2019}.

\begin{figure}[hbt!]
  \centering
\includegraphics[width=0.99\columnwidth]{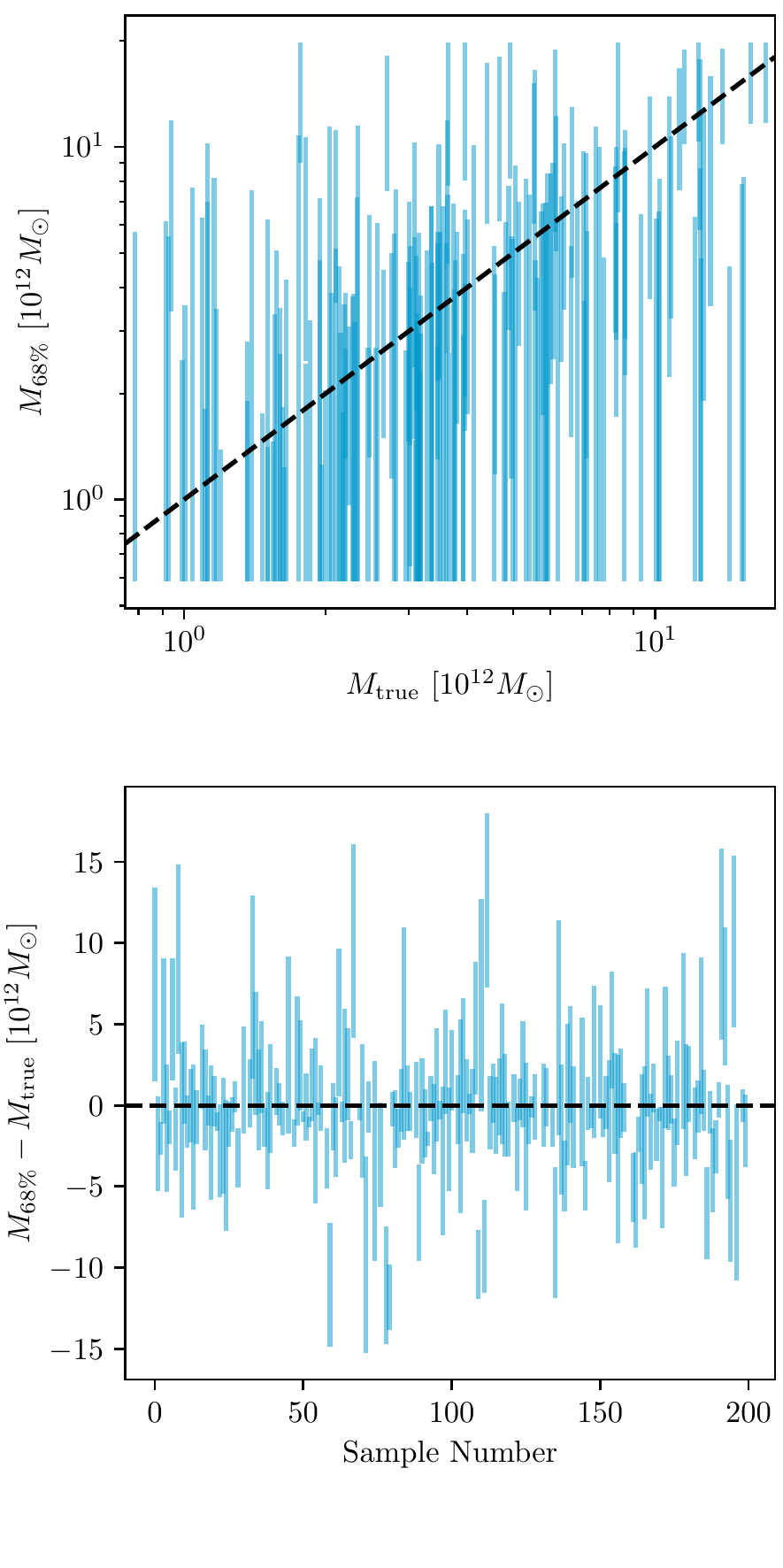}
  \caption{
Validation plot for our density estimation. 
We use 10,000 simulations that 
have not be used for training. For 
all these, we use $r$, $v_r$ and $v_t$ to 
estimate a mass, and compare with the true mass. The top figure plots predicted vs
true mass, while the bottom plot shows
the residuals. The bars show the 68 \% CL obtained using the method described in this paper. 
\label{fig:validation}
}
\end{figure}

We hold back 10,000 simulations from the training set to be used for validating 
the likelihood that we have learned through the simulations.
Each validation simulation has a `true' mass, position and velocity, 
and we can use these to estimate how well our likelihood works. The results from this validation are shown in \cref{fig:validation}.
We also perform a quartile test, finding that $95.485 \%$ of the simulations fall within the $2 \sigma$ predicted posterior,
as expected. 

\subsection{Prior distribution}

As previously discussed, we have the freedom to choose a suitable prior distribution (this is because we have used the
simulations to learn a likelihood function, instead of directly learning the posterior distribution).
The left panel of \cref{fig:priors} shows four priors relevant to this study.
Uninformative priors in this case could be either a flat prior or a logarithmic prior.
In addition, in this problem Press-Schechter theory \cite{Press1974} supplies us with a physically-motivated prior.
The Press-Schechter formalism predicts the number of
virialized objects with a given mass. While this would be a fully correct prior only if the MW and M31 formed a single halo, it can provide a good prior distribution for the problem\footnote{
The Local Group is a bound system but not a virialized system, so placing it in the Press-Schechter mass function 
works as an approximation.
}. We calculate the Press-Schechter prior using the code {\tt Colossus} 
\citep{Diemer:2017bwl}\footnote{\url{https://bdiemer.bitbucket.io/colossus/index.html}}, and the Tinker mass function \citep{Tinker:2008ff}. Finally, the prior shown as a black 
dashed line is the one that would have been in use if we have learned the posterior directly from the simulations (when learning
the posterior directly, we still have a prior, we simply lose the freedom to choose it). 
In this work, we adopt the Press-Schechter prior, which as shown in \cref{fig:priors} is 
virtually equivalent to a flat prior in $\log M$. The effect of using different priors 
in our results will be discussed in \cref{sec:priors}.

Once we have obtained a likelihood function and prior, we can get a posterior using Bayes' theorem \cref{eq:bayes}.
We can describe this posterior by sampling from it using an
algorithm such as Markov chain Monte Carlo (MCMC)
or Nested Sampling \citep{skilling2006}.
However, in our case, a `brute-force' approach is more practical (because the posterior is only one-dimensional):
we simply calculate the posterior on a grid of mass values. 

\begin{figure}
  \centering
    \includegraphics[width=0.99\columnwidth]{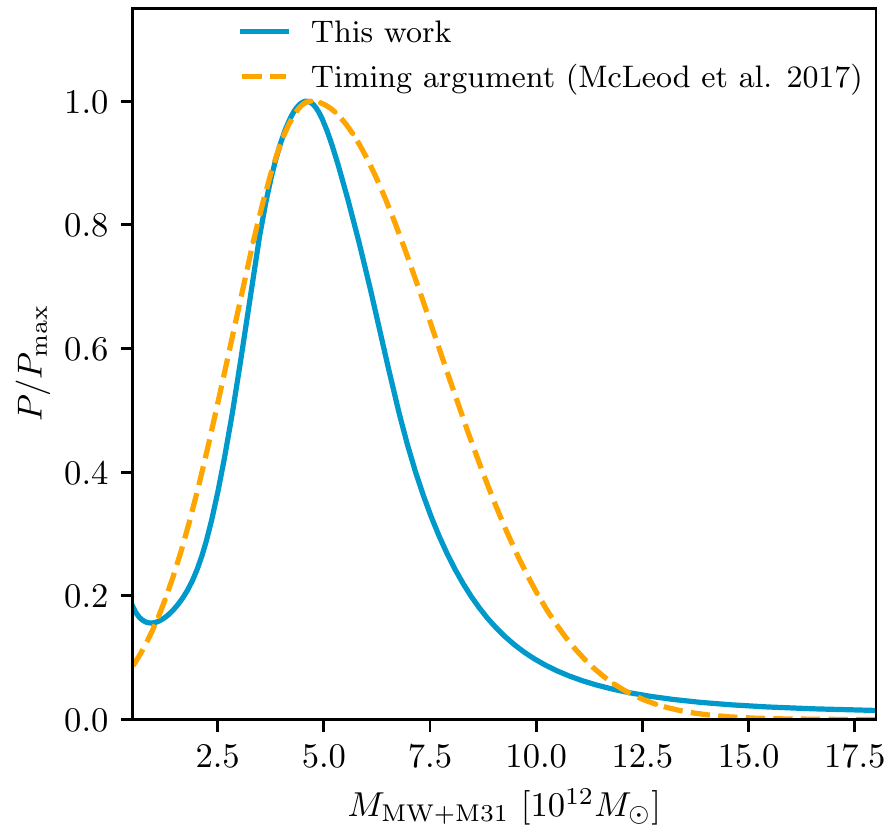}
  \caption{
The posterior on \Mlg{} obtained in this work (solid blue) compared to the timing argument result 
of \citep{McLeod2017} that includes $\Lambda$ and the tangential velocity from \citep{vdm2019} (dashed orange). While our peak is lower, 
the posteriors are fully consistent.
\label{fig:posterior}
}
\end{figure}

Our result using the Press-Schechter prior is shown in \cref{fig:posterior} (solid blue). Our peak and $68\%$ confidence levels are 
$M_{\rm MW+M31} = $ \SI[parse-numbers=false]{4.6^{+2.3}_{-1.8} \times 10^{12}}{\solarmass},
in good agreement with \citep{McLeod2017} (also shown in \cref{fig:posterior}) but with 
improved error bars. 

\subsection{Results discussion}

\begin{figure}[h!]
  \centering
    \includegraphics[width=0.99\columnwidth]{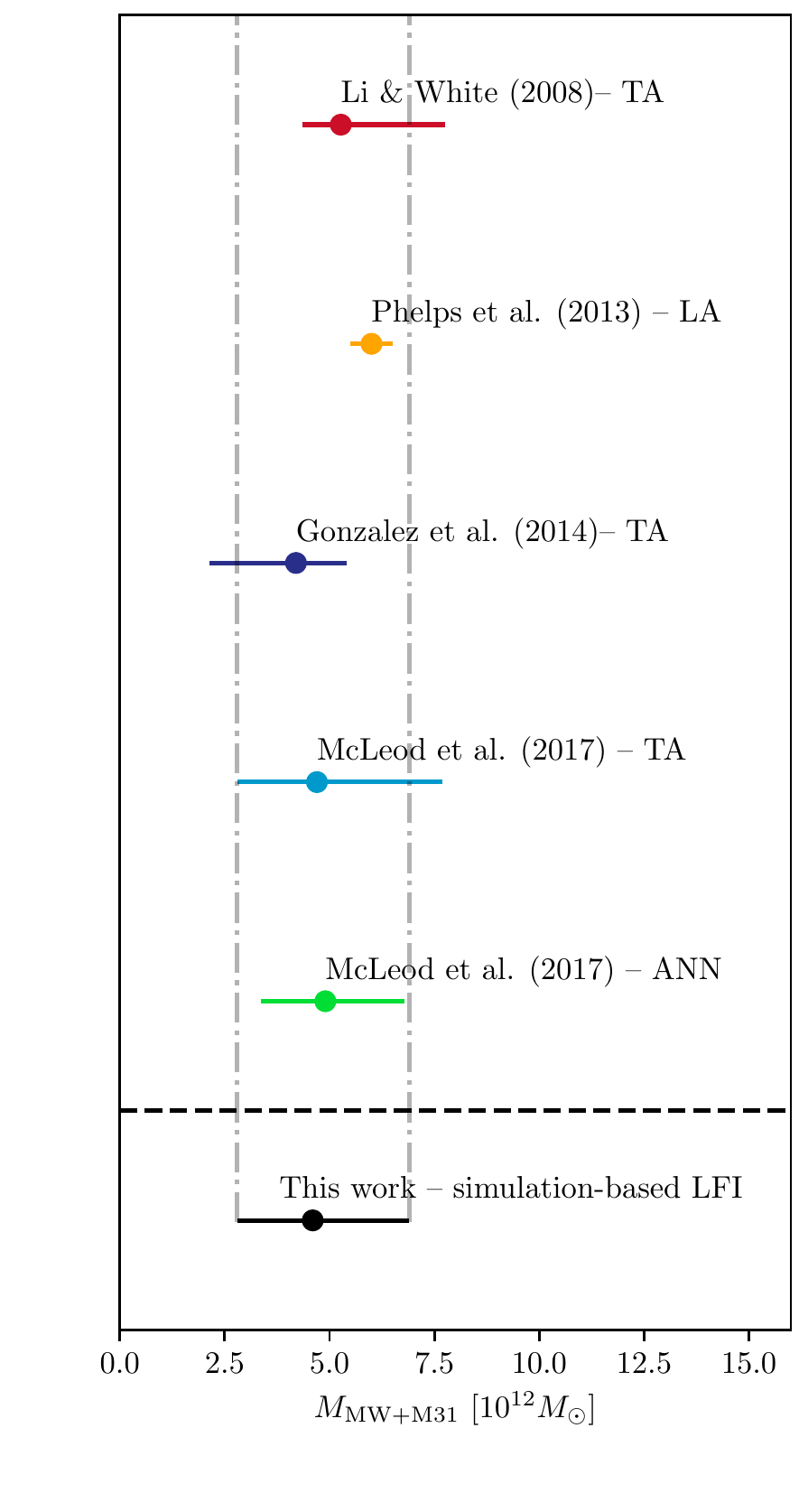}
  \caption{
Comparison of this work with previous estimates of the \Mlg{}, shown as best fit and $68 \%$ confidence 
intervals. The result of this work is shown at the bottom; it
is the first to account fully for the observational errors, and
to not rely on the approximation of the TA.
\label{fig:compare} 
}
\end{figure}

Our result is compared to previous results in \cref{fig:compare}. We see that 
all other estimates considered in this work are within the $68 \%$ confidence interval
of our posterior in the mass, despite the different methods used. We notice that the 
Least Action result
of \citep{Phelps:2013rra}) obtains tighter constraints than our method; however, our result is the first one 
to fully account for the distribution of the observed errors in a robust (and Bayesian) manner. Other results use 
Gaussian approximations for observational
errors, or neglect them completely, and therefore our result is the most accurate estimate
of \Mlg{} to date. This framework also allows for more accurate estimates, 
in particular accounting for the presence of M33 and the LMC in the local group. This will be explored in future work.

The simulation was run using one particular set of cosmological parameters
but in reality these parameters are uncertain and we should marginalise over
them. This is infeasible for us, as we have a pre-run set of simulations with fixed cosmological parameters, but we can estimate the size of the
effect by reference to the timing argument (TA).

The TA uses the same
observational constraints as does this work, and like this work is based on
modelling/simulating the trajectories of galaxies similar to those in the
MW+M31 system; as a result the TA should have similar sensitivities to cosmological
parameters as this work. The TA sensitivities can be estimated by numerically
differentiating the mass estimation algorithm described in
\cite{Partridge:2013dsa}. 
We parameterise this algorithm using $h$ and $\Omega_{\Lambda}$
(from which $\Lambda$ and the age of the universe may be derived,
the latter assuming $\Omega_m + \Omega_{\Lambda} = 1$).
We find
$\partial M_{\rm MW+M31} / \partial \Omega_{\Lambda} = \SI{-2.4E12}{\solarmass}$ and
$\partial M_{\rm MW+M31} / \partial h = \SI{7.4E12}{\solarmass}$.
Multiplying these sensitivities by uncertainties on cosmological parameters
($\Delta \Omega_{\Lambda} = 0.006$ and $\Delta h = 0.004$ \cite{planck_2018_VI})
yields uncertainties on the mass estimate that are immaterial compared
to the uncertainty implied by the posterior width, and hence will be ignored.
This conclusion continues to hold even if we assume a larger uncertainty on $h$ reflecting
the current tension between early- and late-Universe measurements of this parameter. For
example, a change in $h$ of
\SI{0.066} 
induces a
change in the TA \Mlg{} of \SI{0.49E12}{\solarmass} (in agreement with \citep{2019arXiv190310849M});
adding this in quadrature to the uncertainty implied by the posterior width yields only
a marginal increase in total uncertainty
(from \SI{2.3E12}{\solarmass} to \SI{2.35E12}{\solarmass}).
This calculation illustrates that in a simulation-based approach it is important to have a
benchmark analytical model, to gauge if parameters not explored by the simulations are relevant
and if extra simulations are needed.

\begin{figure}[h!]
  \centering
    \includegraphics[width=0.99\columnwidth]{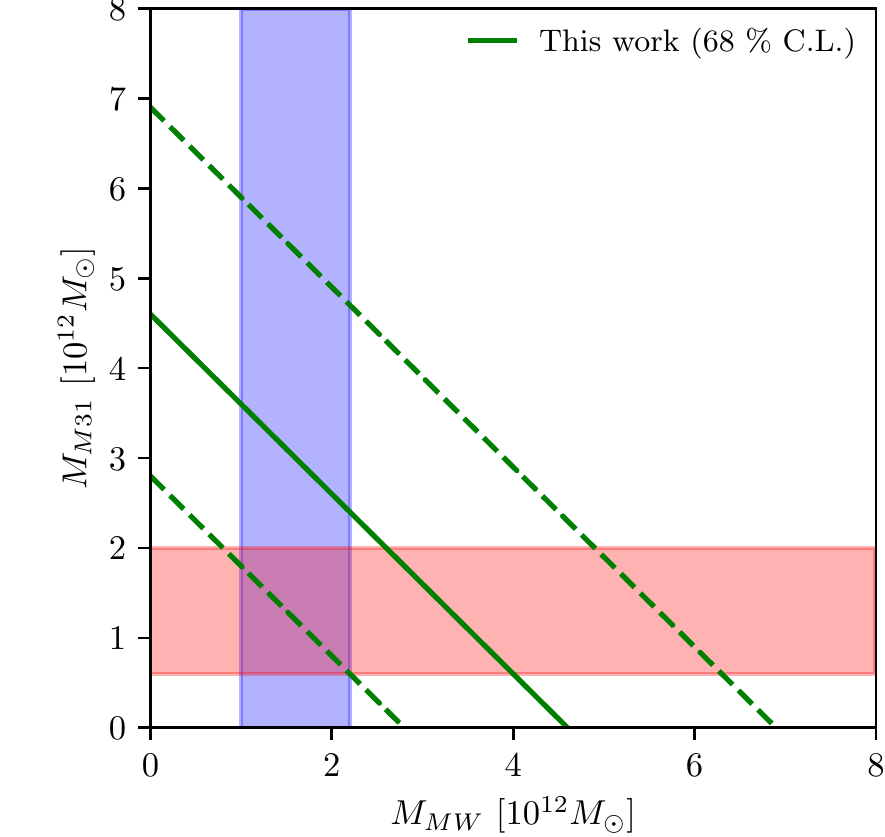}
  \caption{
A comparison of the estimates of the separate
masses of M31, the MW, and their sum, the 
latter from this work. The plot shows the
small discrepancy between separate estimates of the
individual masses of the MW and M31 and this work.
\label{fig:separate}
}
\end{figure}

Finally, we can compare our results with separate estimates of the masses of the
Milky Way and M31. There are several values in the literature for the separate
masses of each galaxy, in some cases discrepant. Given this discrepancy, we take a 
number of estimates of each mass obtained through different methods, 
and assume that the true value is contained within the ranges of the different
estimates, as done in \cite{2020MNRAS.498.2968L}. While conservative, this method should provide with ranges that contain 
the true mass of each galaxy, and allow us to combine estimates in tension.
Through this method, 
we get the following:

\begin{itemize}
\item $M_{\rm MW} \in (1.0,2.2)$ \SI{}{\times 10^{12} \solarmass}, from 
\citep{Diaz2014, Zaritsky2017, Hattori:2018isv, Posti2019, Watkins2019, Karukes:2019jwa} 
\item $M_{\rm M31} \in (0.6,2.0)$ \SI{}{\times 10^{12} \solarmass}, from 
\citep{Corbelli2010, Tamm2012, Diaz2014, Kafle:2018amm}
\end{itemize}

Combining these two measurements yields
$M_{\rm MW + M31} \in (1.6, 4.2)$ \SI{}{\times 10^{12} \solarmass}.
This is slightly lower than our result, but still in agreement, as illustrated in \cref{fig:separate}.
We can see in \cref{fig:compare} that all estimates of the sum of the masses based on the relative distance
and velocity of the bodies (TA, ANN and our approach) obtain slightly larger values than the sum of the separate 
masses. A possible explanation for this could be the fact that all these approaches ignore the effect of other bodies
such as the LMC and M33 in the observed velocities, which could bias the sum of the masses to higher values~\citep{2016MNRAS.456L..54P}. 
The effect of the LMC and M33 in our posterior mass will be explored in future work.

\section{Conclusions}
\label{sec:conclusions}

In this work we have used Density Estimation Likelihood-Free Inference with forward-modelling
to estimate the posterior distribution for sum of the masses of the Milky Way and M31
using observations of the distance and velocity to M31. We obtain a mass
$M_{\rm MW+M31} =$ \SI[parse-numbers=false]{4.6^{+2.3}_{-1.8} \times 10^{12}}{\solarmass} ($M_{200}$).
Our method overcomes the several approximations of the 
traditional Timing Argument, accounts for non-Gaussian sources of observational measurement error, and 
uses a physically motivated prior; 
this makes it the most reliable estimate of \Mlg{} mass to date. 


The sensitivity analysis performed in this study illustrates that in any simulation-based approach
it is important to have a benchmark analytical (or semi-analytical) model, to assess how to cover
the parameter space of required simulations.

This works serves not only to obtain state-of-the-art estimates of the \Mlg{}; by applying Likelihood-Free
Inference to a problem that is physically rich and complex yet statistically simple 
(thanks to its low dimensionality), we can illustrate how the method works, what different choices need to be 
made, and what challenges need to be tackled. The ability to robustly infer \Mlg{} without requiring an analytic 
theory or a likelihood demonstrates the potential of Likelihood-Free Inference methods in astronomy and cosmology.

\begin{acknowledgements}
We thank Andrey Kravtsov, Roeland van der Marel, Ekta Patel and Vasily Belokurov. We are also grateful to David Benisty for participating in discussions that led to the idea of this paper.

PL \& OL acknowledge STFC Consolidated Grant ST/R000476/1.
NIL acknowledges financial support of the Project IDEXLYON at the University of Lyon under the Investments for the Future Program (ANR-16-IDEX-0005).
YH has been partially supported by the Israel Science Foundation grant ISF 1358/18.
\end{acknowledgements}

\appendix

\section{Density Estimators}
\label{sec:density}

\begin{figure*}

  \hspace*{-1cm} 
     \includegraphics[width=1.99\columnwidth]{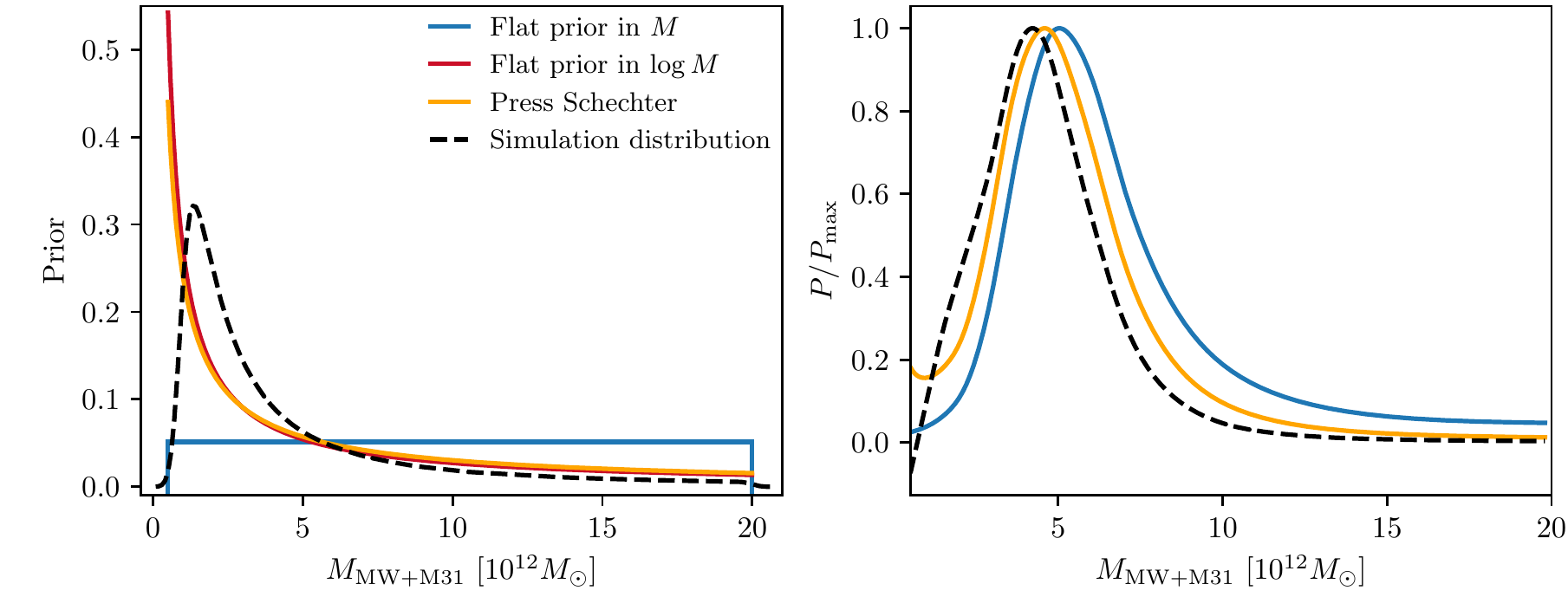}
  \caption{
On the left, a comparison of four possible priors: a flat prior in the mass (blue), a flat prior in the logarithm of the 
mass (red), a physically motivated Press-Schechter prior (orange), and the prior from the distribution of the simulations (dashed black). 
On the right, the corresponding posterior
obtained from each prior. Note that
we did not show the posterior for the flat prior in the logarithm of the 
mass, as this is equivalent to 
the Press-Schechter prior. 
\label{fig:priors}
}
\end{figure*}

One of the key elements of DELFI is the estimation of a probability distribution from samples.
This corresponds to going from \cref{fig:delfi} (right panel) to \cref{fig:delfi2}. 
The density estimation problem arises in many fields (for example image analysis
\citep[][]{theis2015generative, Salimans2017})
and several algorithms have been developed to address it.
In this section, we review some of the most popular density estimation methods in the context of LFI. For an overview of neural density estimation in the context of LFI we recommend~\cite{Alsing2019}.

Density estimation algorithms that rely on there being samples near the point of interest, such as spline or
Kernel Density Estimation (KDE), struggle in high dimensional spaces due to the sparsity of the sampling. They are
very useful, however, for estimating low dimensional PDFs, which is why they are often used for plotting marginalised
posterior distributions. Public codes such as {\tt GetDist} \citep{Lewis:2019xzd}, {\tt ChainConsumer} 
\citep{Hinton2016} or {\tt anesthetic} \citep{anesthetic} use KDE to 
generate plots of marginalised posterior distributions.

A Mixture Model (MM) represents a PDF $p$ as a weighted sum of component distributions:
\begin{equation}
\label{eq:MixtureModel}
p(\mathbf{y} ) = \sum_{c=1}^N \alpha_c \mathcal{D} (\mathbf{y}; \Phi_c).
\end{equation}
Here $N$ is the number of components in the mixture while $\mathcal{D}$ is some family of distributions described
by parameters $\Phi$; the weights $\left\{ \alpha_c \right\}$ and parameters $\left\{ \Phi_c \right\}$ are fit to
observed or training data. A common choice is the Gaussian Mixture Model (GMM), in which each component distribution
is Gaussian: $\mathcal{D}(\mathbf{y}; \Phi_c) = \mathcal{N} ( \mathbf{y}; \mu_c, \sigma_c )$. 

GMMs can successfully represent a large number of PDFs. In addition, they have the advantage that the weights and
parameters $\left\{ \alpha_c, \mu_c, \sigma_c \right\}$ can be easily fit to the data using the 
Expectation-Maximization algorithm \citep{Dempster1977}. There are, however, some issues with GMMs: they are 
sensitive to the choice of $N$, and they have problems fitting certain features (such as the sharp edges that
can arise when flat priors are used).

In the context of LFI we are interested in modelling a \textit{conditional} distribution $p(\mathbf{y}|\mathbf{x})$
(for example in our case $p$ is the conditional likelihood, for which $\mathbf{y}=d$ and $\mathbf{x}=\theta$).
Such conditional distributions can be modelled by Mixture Density Networks (MDNs)
\citep{Bishop94mixturedensity, Bishop2006}.
As with MMs, they model the PDF as a weighted sum of component distributions, but now the
weights and parameters describing the components are themselves (possibly non-linear) functions of $\mathbf{x}$:
\begin{equation}
\label{eq:MDN}
p(\mathbf{y} | \mathbf{x}) = \sum_{c=1}^N \mathbf{\alpha}_c (\mathbf{x}) \mathcal{D}_c (\mathbf{y} ; \Phi_c (\mathbf{x})).
\end{equation}
Again, a common choice is the Gaussian MDN (GMDN), in which each component is Gaussian:
$\mathcal{D}(\mathbf{y}; \Phi_c (\mathbf{x})) = \mathcal{N} ( \mathbf{y}; \mu_c(\mathbf{x}), \sigma_c(\mathbf{x}) )$.

The functions $\left\{ \mathbf{\alpha}_c (\mathbf{x}), \mu_c (\mathbf{x}), \sigma_c (\mathbf{x}) \right\}$ can be 
modelled by a neural network with a set of weights; these weights are then fit to the data. As with GMMs, GMDNs
require specification of the number $N$ of mixture components to be used; however, this dependence is much smaller
than in the case of GMMs (as GMDNs can fit complex distributions using only a small number of components). 

We finish by describing Masked Autoregressive Flows (MAFs), which have recently emerged as a powerful density
estimation method \citep{Papamakarios2017, papamakarios2019sequential}. They do not rely on a choice of number of components,
and have the advantage of providing simple tests of the goodness of fit to the samples. 

Here is the motivation for \textit{masking} as a strategy for density estimation. Consider training a neural network
\textit{NN} to mimic; one can imagine training a parrot, for example. The trainer speaks (= input signal), and
rewards the bird if its output matches this training input. If \textit{NN} has sufficient complexity then it will learn to
mimic the input. Now repeat the process but with a bird with covered (= masked) ears. The bird can't hear the input,
but nevertheless receives the training reward if its output matches the input. Now the bird can only `play the 
percentages'; it learns the optimal strategy, which is to output a weighted average of the input signals 
(weighted by their frequency of usage by the trainer). In this way the masked parrot learns the probability 
distribution of the input signal i.e. has become a density estimator.

That was the one-dimensional case. The two dimensional case needs two parrots. The first is trained on signal
$x_1$, which it can't hear, with the result that it learns $p(x_1)$. The second is trained on $x_2$, which it can't
hear, but it is allowed to hear $x_1$. As a result it learns $p(x_2|x_1)$. Thus between them they learn
$p(x_1) p(x_2|x_1) = p(x_1, x_2)$ as desired. The multi-dimensional case is similar. This strategy is called an
\textit{autoregressive autoencoder}. Note that it treats the coordinates asymmetrically.

The conditional distributions $p(x_i|x_1, \dots, x_{i-1})$ learned by \textit{NN} are typically modelled as Gaussian.
Consider generating samples from the estimated probability distribution $p$; for each sample we need we a set of
$n$ random unit normals (i.e. a draw from $N(0,I)$), which we transform to get samples from $p$ - call this
transform $T$. The details of $T$ come from the means and standard deviations of the conditional distributions,
which can be obtained from $n$ evaluations of \textit{NN}. However, importantly, the inverse mapping $T^{-1}$ can be
found with just \textit{one} evaluation of \textit{NN}. This is the idea of the Masked Autoencoder for Distribution
Estimation (MADE) algorithm \citep{germain2015made}.

Now apply $T^{-1}$ to the training data $D$. The resulting `pulled-back' data $T^{-1}(D)$ will ideally be a
set of samples from $N(0, I)$ and its deviation from this ideal gives a direct measure of how imperfect is our
modelling of $p$. We can then use the `pulled-back' data as the training data for yet another MADE
process, and so on through several iterations. Between iterations we permute the coordinate axes,
thereby symmetrising how we treat them. With sufficient iterations, the multiply-pulled-back training data
approaches $N(0,I)$; at this point the algorithm has an easy-to-evaluate mapping between $p$ and $N(0, I)$, which
suffices for doing calculations. This is the Masked Autoregressive Flows (MAF) algorithm
\citep{Papamakarios2017, papamakarios2019sequential}.

\section{Dependence on Priors}
\label{sec:priors}

In this appendix, we explore how the posterior 
distribution of \Mlg{} (as shown in \cref{fig:posterior} and discussed in 
\cref{sec:results}) depends on our choice of prior.
We consider the four different priors illustrated in \cref{fig:priors}:

\begin{itemize}
    \item A flat prior in the mass;
    \item A logarithmic prior in the mass;
    \item A prior based on the Press-Schechter distribution (as adopted in this work);
    \item A prior distribution matching the distribution of masses in the simulation.
\end{itemize}

In our case the second and third choices are virtually the same (as shown in the left panel of \cref{fig:priors}), 
and so we omit the `logarithmic in mass' prior when examining how the priors affect our results. The 
posteriors obtained when using the remaining three priors are shown in the right panel of \cref{fig:priors},
and we see that our result is essentially independent of our choice of prior, be it the Press-Schechter prior, 
a flat prior on the mass
($M_{\rm LG} (\textrm{Flat prior}) = 5.0^{+2.7}_{-1.7} \times 10^{12} M_{\odot}$) 
or the prior from the simulation distribution.
($M_{\rm LG} (\textrm{Simulation Distribution}) = 4.3 \pm 1.7 \times 10^{12} M_{\odot}$)

\section{Dependence on Tangential Velocity}
\label{sec:vt}

\begin{figure}
  \centering
    \includegraphics[width=0.99\columnwidth]{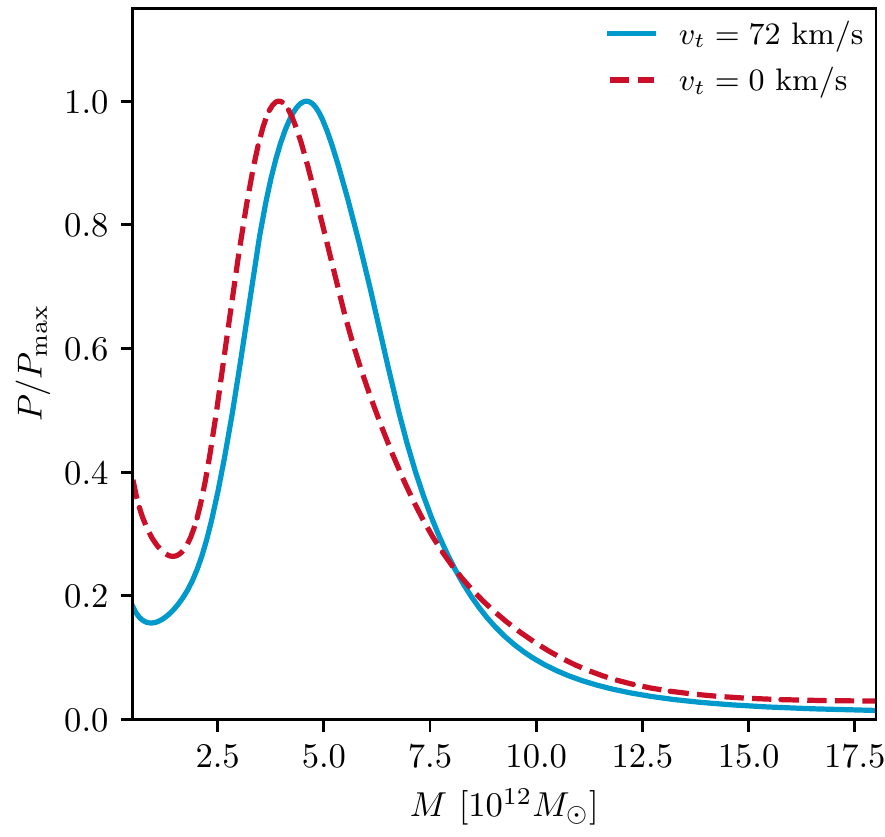}
  \caption{
The posterior on \Mlg{} for different values of the 
tangential velocity: $v_t =$ \SI{72}{\kilo\metre\per\s} as used in this work (solid blue),
and a 
purely radial motion $v_t = 0$ (dashed red).
\label{fig:vt_posterior}
}

\end{figure}


Imagine a 2-dimensional velocity vector ${\bf V} = (V_x,V_y)$.
If $V_x,V_y$ are uncorrelated, normally distributed with zero mean and equal variance $\sigma$, then the overall  speed $V={\sqrt {(V_x^2+V_y^2)}}$ will be characterized by the Rayleigh distribution \citep{rayleigh1905problem}, with mean ${\bar V}= \sigma \sqrt {\pi/2}$, rather than naively zero. Similarly, non-zero measurements of ${V_x,V_y}$ with error bars will result in a distribution function with a mean that is {\it not} ${\bar V}={\sqrt {(V_x^2+V_y^2)}}$.
In our analysis we pay attention to this via the forward modeling approach, starting with simulated $(V_x, V_y)$ and propagating their impact on the final posterior for \Mlg{}.

While distance and radial velocity
have been measured in numerous occasions using different methods, observations of the tangential velocity are far
more scarce \citep{vanderMarel:2007yw, vdm2012, vdm2019}. Their work treats the effect of converting
measurements of two components of the tangential velocity into a modulus in a novel way. 
We check the robustness of our approach in this section. We do 
so by comparing the main result of the paper to the case of no tangential velocity. As shown in \cref{fig:vt_posterior}, 
our posterior on the mass does not depend strongly on the tangential velocity. Therefore, we are confident on 
the accuracy of our posterior in the mass. 

\bibliographystyle{unsrtnat}
\bibliography{LG}

\begin{thebibliography}{81}
\providecommand{\natexlab}[1]{#1}
\providecommand{\url}[1]{\texttt{#1}}
\expandafter\ifx\csname urlstyle\endcsname\relax
  \providecommand{\doi}[1]{doi: #1}\else
  \providecommand{\doi}{doi: \begingroup \urlstyle{rm}\Url}\fi

\bibitem[{Leclercq}(2018)]{Leclercq2018}
Florent {Leclercq}.
\newblock {Bayesian optimization for likelihood-free cosmological inference}.
\newblock \emph{\prd}, 98\penalty0 (6):\penalty0 063511, September 2018.
\newblock \doi{10.1103/PhysRevD.98.063511}.

\bibitem[{Alsing} et~al.(2019){Alsing}, {Charnock}, {Feeney}, and {Wand
  elt}]{Alsing2019}
Justin {Alsing}, Tom {Charnock}, Stephen {Feeney}, and Benjamin {Wand elt}.
\newblock {Fast likelihood-free cosmology with neural density estimators and
  active learning}.
\newblock \emph{\mnras}, 488\penalty0 (3):\penalty0 4440--4458, September 2019.
\newblock \doi{10.1093/mnras/stz1960}.

\bibitem[{Betoule} et~al.(2014){Betoule}, {Kessler}, {Guy}, {Mosher}, and
  {Hardin et al.}]{Betoule2014}
M.~{Betoule}, R.~{Kessler}, J.~{Guy}, J.~{Mosher}, and D.~{Hardin et al.}
\newblock {Improved cosmological constraints from a joint analysis of the
  SDSS-II and SNLS supernova samples}.
\newblock \emph{\aap}, 568:\penalty0 A22, August 2014.
\newblock \doi{10.1051/0004-6361/201423413}.

\bibitem[{Wang} et~al.(2020){Wang}, {Xie}, {Zhang}, {Huang}, {Zhang}, and
  {Liu}]{2020arXiv200510628W}
Yu-Chen {Wang}, Yuan-Bo {Xie}, Tong-Jie {Zhang}, Hui-Chao {Huang}, Tingting
  {Zhang}, and Kun {Liu}.
\newblock {Likelihood-free Cosmological Constraints with Artificial Neural
  Networks: An Application on Hubble Parameters and SN Ia}.
\newblock \emph{arXiv e-prints}, art. arXiv:2005.10628, May 2020.

\bibitem[{Jeffrey} et~al.(2020){Jeffrey}, {Alsing}, and {Lanusse}]{Jeffrey2020}
Niall {Jeffrey}, Justin {Alsing}, and Francois {Lanusse}.
\newblock {Likelihood-free inference with neural compression of DES SV weak
  lensing map statistics}.
\newblock \emph{arXiv e-prints}, art. arXiv:2009.08459, September 2020.

\bibitem[Brehmer et~al.(2019)Brehmer, Mishra-Sharma, Hermans, Louppe, and
  Cranmer]{Brehmer:2019jyt}
Johann Brehmer, Siddharth Mishra-Sharma, Joeri Hermans, Gilles Louppe, and Kyle
  Cranmer.
\newblock {Mining for Dark Matter Substructure: Inferring subhalo population
  properties from strong lenses with machine learning}.
\newblock 9 2019.
\newblock \doi{10.3847/1538-4357/ab4c41}.

\bibitem[Ramanah et~al.(2020)Ramanah, Wojtak, and Arendse]{Ramanah:2020ylz}
Doogesh~Kodi Ramanah, Rados\l{}aw Wojtak, and Nikki Arendse.
\newblock {Simulation-based inference of dynamical galaxy cluster masses with
  3D convolutional neural networks}.
\newblock 9 2020.

\bibitem[{Tortorelli} et~al.(2020){Tortorelli}, {Fagioli}, {Herbel}, {Amara},
  {Kacprzak}, and {Refregier}]{2020JCAP...09..048T}
Luca {Tortorelli}, Martina {Fagioli}, J{\"o}rg {Herbel}, Adam {Amara}, Tomasz
  {Kacprzak}, and Alexandre {Refregier}.
\newblock {Measurement of the B-band galaxy Luminosity Function with
  Approximate Bayesian Computation}.
\newblock \emph{\jcap}, 2020\penalty0 (9):\penalty0 048, September 2020.
\newblock \doi{10.1088/1475-7516/2020/09/048}.

\bibitem[{Kahn} and {Woltjer}(1959)]{Kahn1959}
F.~D. {Kahn} and L.~{Woltjer}.
\newblock {Intergalactic Matter and the Galaxy.}
\newblock \emph{\apj}, 130:\penalty0 705, November 1959.
\newblock \doi{10.1086/146762}.

\bibitem[{Prada} et~al.(2012){Prada}, {Klypin}, {Cuesta}, {Betancort-Rijo}, and
  {Primack}]{Prada2012}
Francisco {Prada}, Anatoly~A. {Klypin}, Antonio~J. {Cuesta}, Juan~E.
  {Betancort-Rijo}, and Joel {Primack}.
\newblock {Halo concentrations in the standard {\ensuremath{\Lambda}} cold dark
  matter cosmology}.
\newblock \emph{\mnras}, 423\penalty0 (4):\penalty0 3018--3030, July 2012.
\newblock \doi{10.1111/j.1365-2966.2012.21007.x}.

\bibitem[{Riebe} et~al.(2013){Riebe}, {Partl}, {Enke}, {Forero-Romero},
  {Gottl{\"o}ber}, {Klypin}, {Lemson}, {Prada}, {Primack}, {Steinmetz}, and
  {Turchaninov}]{2013AN....334..691R}
K.~{Riebe}, A.~M. {Partl}, H.~{Enke}, J.~{Forero-Romero}, S.~{Gottl{\"o}ber},
  A.~{Klypin}, G.~{Lemson}, F.~{Prada}, J.~R. {Primack}, M.~{Steinmetz}, and
  V.~{Turchaninov}.
\newblock {The MultiDark Database: Release of the Bolshoi and MultiDark
  cosmological simulations}.
\newblock \emph{Astronomische Nachrichten}, 334\penalty0 (7):\penalty0
  691--708, August 2013.
\newblock \doi{10.1002/asna.201211900}.

\bibitem[{Meylan} et~al.(2004){Meylan}, {Madrid}, and {Macchetto}]{HST}
Georges {Meylan}, Juan~P. {Madrid}, and Duccio {Macchetto}.
\newblock {Hubble Space Telescope Science Metrics}.
\newblock \emph{\pasp}, 116\penalty0 (822):\penalty0 790--796, August 2004.
\newblock \doi{10.1086/423227}.

\bibitem[{Gaia Collaboration}(2016)]{Gaia}
{Gaia Collaboration}.
\newblock {The Gaia mission}.
\newblock \emph{\aap}, 595:\penalty0 A1, November 2016.
\newblock \doi{10.1051/0004-6361/201629272}.

\bibitem[{McLeod} et~al.(2017){McLeod}, {Libeskind}, {Lahav}, and
  {Hoffman}]{McLeod2017}
M.~{McLeod}, N.~{Libeskind}, O.~{Lahav}, and Y.~{Hoffman}.
\newblock {Estimating the mass of the Local Group using machine learning
  applied to numerical simulations}.
\newblock \emph{\jcap}, 2017\penalty0 (12):\penalty0 034, December 2017.
\newblock \doi{10.1088/1475-7516/2017/12/034}.

\bibitem[{Bonassi} et~al.(2011){Bonassi}, {You}, and {West}]{Bonassi2011}
F.V. {Bonassi}, L.~{You}, and M.~{West}.
\newblock {Bayesian Learning from Marginal Data in Bionetwork Models}.
\newblock \emph{Statistical applications in genetics and molecular biology},
  2011\penalty0 (10,1):\penalty0 49, October 2011.
\newblock \doi{10.2202/1544-6115.1684}.

\bibitem[{Fan} et~al.(2012){Fan}, {Nott}, and {Sisson}]{Fan2012}
Y.~{Fan}, D.~J. {Nott}, and S.~A. {Sisson}.
\newblock {Approximate Bayesian Computation via Regression Density Estimation}.
\newblock \emph{arXiv e-prints}, art. arXiv:1212.1479, December 2012.

\bibitem[{Papamakarios} and {Murray}(2016)]{Papamakarios2016}
George {Papamakarios} and Iain {Murray}.
\newblock {Fast $\epsilon$-free Inference of Simulation Models with Bayesian
  Conditional Density Estimation}.
\newblock \emph{arXiv e-prints}, art. arXiv:1605.06376, May 2016.

\bibitem[Li and White(2008)]{Li:2007eg}
Yang-Shyang Li and Simon~D.M. White.
\newblock {Masses for the Local Group and the Milky Way}.
\newblock \emph{Mon. Not. Roy. Astron. Soc.}, 384:\penalty0 1459--1468, 2008.
\newblock \doi{10.1111/j.1365-2966.2007.12748.x}.

\bibitem[Gonzalez et~al.(2014)Gonzalez, Kravtsov, and Gnedin]{Gonzalez:2013pqa}
Roberto~E. Gonzalez, Andrey~V. Kravtsov, and Nickolay~Y. Gnedin.
\newblock {On the mass of the Local Group}.
\newblock \emph{Astrophys. J.}, 793:\penalty0 91, 2014.
\newblock \doi{10.1088/0004-637X/793/2/91}.

\bibitem[Phelps et~al.(2013)Phelps, Nusser, and Desjacques]{Phelps:2013rra}
Steven Phelps, Adi Nusser, and Vincent Desjacques.
\newblock {The mass of the Milky Way and M31 using the method of least action}.
\newblock \emph{Astrophys. J.}, 775:\penalty0 102, 2013.
\newblock \doi{10.1088/0004-637X/775/2/102}.

\bibitem[{Lynden-Bell}(1981)]{Lynden-Bell1981}
D.~{Lynden-Bell}.
\newblock {The dynamical age of the local group of galaxies}.
\newblock \emph{The Observatory}, 101:\penalty0 111--114, August 1981.

\bibitem[{Binney} and {Tremaine}(1987)]{Binney1987}
James {Binney} and Scott {Tremaine}.
\newblock \emph{{Galactic dynamics}}.
\newblock 1987.

\bibitem[Partridge et~al.(2013)Partridge, Lahav, and
  Hoffman]{Partridge:2013dsa}
Candace Partridge, Ofer Lahav, and Yehuda Hoffman.
\newblock {Weighing the Local Group in the Presence of Dark Energy}.
\newblock \emph{Mon. Not. Roy. Astron. Soc.}, 436:\penalty0 45, 2013.
\newblock \doi{10.1093/mnrasl/slt109}.

\bibitem[{McLeod} and {Lahav}(2019)]{2019arXiv190310849M}
Michael {McLeod} and Ofer {Lahav}.
\newblock {The Two Body Problem in the Presence of Dark Energy and Modified
  Gravity: Application to the Local Group}.
\newblock \emph{arXiv e-prints}, art. arXiv:1903.10849, March 2019.

\bibitem[{Benisty} et~al.(2019){Benisty}, {Guendelman}, and
  {Lahav}]{2019arXiv190403153B}
David {Benisty}, Eduardo~I. {Guendelman}, and Ofer {Lahav}.
\newblock {Milky Way and Andromeda past-encounters in different gravity models:
  the impact on the estimated Local Group mass}.
\newblock \emph{arXiv e-prints}, art. arXiv:1904.03153, April 2019.

\bibitem[{Peebles}(1994)]{Peebles1994}
P.~J.~E. {Peebles}.
\newblock {Orbits of the Nearby Galaxies}.
\newblock \emph{\apj}, 429:\penalty0 43, July 1994.
\newblock \doi{10.1086/174301}.

\bibitem[Rubin(1984)]{Rubin1984}
Donald~B Rubin.
\newblock Bayesianly justifiable and relevant frequency calculations for the
  applied statistician.
\newblock \emph{The Annals of Statistics}, pages 1151--1172, 1984.

\bibitem[Rosenblatt()]{rosenblatt1956}
Murray Rosenblatt.
\newblock Remarks on some nonparametric estimates of a density function.
\newblock \emph{Ann. Math. Statist.}, \penalty0 (3):\penalty0 832--837, 09 .
\newblock \doi{10.1214/aoms/1177728190}.

\bibitem[Parzen()]{parzen1962}
Emanuel Parzen.
\newblock On estimation of a probability density function and mode.
\newblock \emph{Ann. Math. Statist.}, \penalty0 (3):\penalty0 1065--1076, 09 .
\newblock \doi{10.1214/aoms/1177704472}.

\bibitem[Simonoff(1996)]{simonoff1996smoothing}
J.S. Simonoff.
\newblock \emph{Smoothing Methods in Statistics}.
\newblock Springer Series in Statistics. Springer, 1996.
\newblock ISBN 9780387947167.
\newblock URL \url{https://books.google.co.uk/books?id=wFTgNXL4feIC}.

\bibitem[Bishop(1994)]{Bishop94mixturedensity}
Christopher~M. Bishop.
\newblock Mixture density networks.
\newblock Technical report, 1994.

\bibitem[Bishop(2006)]{Bishop2006}
Christopher~M. Bishop.
\newblock \emph{Pattern Recognition and Machine Learning (Information Science
  and Statistics)}.
\newblock Springer-Verlag, Berlin, Heidelberg, 2006.
\newblock ISBN 0387310738.

\bibitem[Papamakarios et~al.(2017)Papamakarios, Pavlakou, and
  Murray]{Papamakarios2017}
George Papamakarios, Theo Pavlakou, and Iain Murray.
\newblock Masked autoregressive flow for density estimation.
\newblock In \emph{Advances in Neural Information Processing Systems}, pages
  2338--2347, 2017.

\bibitem[{Cameron} and {Pettitt}(2012)]{Cameron2012}
E.~{Cameron} and A.~N. {Pettitt}.
\newblock {Approximate Bayesian Computation for astronomical model analysis: a
  case study in galaxy demographics and morphological transformation at high
  redshift}.
\newblock \emph{\mnras}, 425\penalty0 (1):\penalty0 44--65, September 2012.
\newblock \doi{10.1111/j.1365-2966.2012.21371.x}.

\bibitem[{Weyant} et~al.(2013){Weyant}, {Schafer}, and
  {Wood-Vasey}]{Weyant2013}
Anja {Weyant}, Chad {Schafer}, and W.~Michael {Wood-Vasey}.
\newblock {Likelihood-free Cosmological Inference with Type Ia Supernovae:
  Approximate Bayesian Computation for a Complete Treatment of Uncertainty}.
\newblock \emph{\apj}, 764\penalty0 (2):\penalty0 116, February 2013.
\newblock \doi{10.1088/0004-637X/764/2/116}.

\bibitem[Akeret et~al.(2015)Akeret, Refregier, Amara, Seehars, and
  Hasner]{Akeret:2015uha}
Joël Akeret, Alexandre Refregier, Adam Amara, Sebastian Seehars, and Caspar
  Hasner.
\newblock {Approximate Bayesian Computation for Forward Modeling in Cosmology}.
\newblock \emph{JCAP}, 08:\penalty0 043, 2015.
\newblock \doi{10.1088/1475-7516/2015/08/043}.

\bibitem[Hahn et~al.(2017)Hahn, Vakili, Walsh, Hearin, Hogg, and
  Campbell]{Hahn:2016zwc}
ChangHoon Hahn, Mohammadjavad Vakili, Kilian Walsh, Andrew~P. Hearin, David~W.
  Hogg, and Duncan Campbell.
\newblock {Approximate Bayesian computation in large-scale structure:
  constraining the galaxy--halo connection}.
\newblock \emph{Mon. Not. Roy. Astron. Soc.}, 469\penalty0 (3):\penalty0
  2791--2805, 2017.
\newblock \doi{10.1093/mnras/stx894}.

\bibitem[Peel et~al.(2017)Peel, Lin, Lanusse, Leonard, Starck, and
  Kilbinger]{Peel:2016jub}
Austin Peel, Chieh-An Lin, Francois Lanusse, Adrienne Leonard, Jean-Luc Starck,
  and Martin Kilbinger.
\newblock {Cosmological constraints with weak lensing peak counts and
  second-order statistics in a large-field survey}.
\newblock \emph{Astron. Astrophys.}, 599:\penalty0 A79, 2017.
\newblock \doi{10.1051/0004-6361/201629928}.

\bibitem[Kacprzak et~al.(2018)Kacprzak, Herbel, Amara, and
  Réfrégier]{Kacprzak:2017nxl}
T.~Kacprzak, J.~Herbel, A.~Amara, and A.~Réfrégier.
\newblock {Accelerating Approximate Bayesian Computation with Quantile
  Regression: application to cosmological redshift distributions}.
\newblock \emph{JCAP}, 02:\penalty0 042, 2018.
\newblock \doi{10.1088/1475-7516/2018/02/042}.

\bibitem[{Alsing} et~al.(2018){Alsing}, {Wandelt}, and {Feeney}]{Alsing2018}
Justin {Alsing}, Benjamin {Wandelt}, and Stephen {Feeney}.
\newblock {Massive optimal data compression and density estimation for
  scalable, likelihood-free inference in cosmology}.
\newblock \emph{\mnras}, 477\penalty0 (3):\penalty0 2874--2885, July 2018.
\newblock \doi{10.1093/mnras/sty819}.

\bibitem[{Alsing} and {Wandelt}(2018)]{Alsing2018b}
Justin {Alsing} and Benjamin {Wandelt}.
\newblock {Generalized massive optimal data compression}.
\newblock \emph{\mnras}, 476\penalty0 (1):\penalty0 L60--L64, May 2018.
\newblock \doi{10.1093/mnrasl/sly029}.

\bibitem[Heavens et~al.(2020)Heavens, Sellentin, and Jaffe]{heavens2020extreme}
Alan~F Heavens, Elena Sellentin, and Andrew~H Jaffe.
\newblock Extreme data compression while searching for new physics.
\newblock \emph{Monthly Notices of the Royal Astronomical Society},
  498\penalty0 (3):\penalty0 3440--3451, 2020.

\bibitem[Papamakarios et~al.(2019)Papamakarios, Sterratt, and
  Murray]{papamakarios2019sequential}
George Papamakarios, David Sterratt, and Iain Murray.
\newblock Sequential neural likelihood: Fast likelihood-free inference with
  autoregressive flows.
\newblock In \emph{The 22nd International Conference on Artificial Intelligence
  and Statistics}, pages 837--848. PMLR, 2019.

\bibitem[Lueckmann et~al.(2019)Lueckmann, Bassetto, Karaletsos, and
  Macke]{lueckmann2019likelihood}
Jan-Matthis Lueckmann, Giacomo Bassetto, Theofanis Karaletsos, and Jakob~H
  Macke.
\newblock Likelihood-free inference with emulator networks.
\newblock In \emph{Symposium on Advances in Approximate Bayesian Inference},
  pages 32--53. PMLR, 2019.

\bibitem[{Knebe} et~al.(2011){Knebe}, {Knollmann}, {Muldrew}, {Pearce}, and
  {Aragon-Calvo et al.}]{Knebe2011}
Alexander {Knebe}, Steffen~R. {Knollmann}, Stuart~I. {Muldrew}, Frazer~R.
  {Pearce}, and Miguel~Angel {Aragon-Calvo et al.}
\newblock {Haloes gone MAD: The Halo-Finder Comparison Project}.
\newblock \emph{\mnras}, 415\penalty0 (3):\penalty0 2293--2318, August 2011.
\newblock \doi{10.1111/j.1365-2966.2011.18858.x}.

\bibitem[Holland(1998)]{Holland:1998br}
Stephen Holland.
\newblock {The Distance to the M31 Globular Cluster System}.
\newblock \emph{Astron. J.}, 115:\penalty0 1916, 1998.
\newblock \doi{10.1086/300348}.

\bibitem[Joshi et~al.(2003)Joshi, Pandey, Narasimha, Sagar, and
  Giraud-Hiraud]{Joshi:2002uf}
Y.C. Joshi, A.K. Pandey, D.~Narasimha, R.~Sagar, and Y.~Giraud-Hiraud.
\newblock {Identification of 13 Cepheids and 333 other variables in M 31}.
\newblock \emph{Astron. Astrophys.}, 402:\penalty0 113--123, 2003.
\newblock \doi{10.1051/0004-6361:20030136}.

\bibitem[Ribas et~al.(2005)Ribas, Jordi, Vilardell, Fitzpatrick, Hilditch, and
  Guinan]{Ribas:2005uw}
Ignasi Ribas, Carme Jordi, Francesc Vilardell, Edward~L. Fitzpatrick, Ron~W.
  Hilditch, and Edward~F. Guinan.
\newblock {First determination of the distance and fundamental properties of an
  eclipsing binary in the andromeda galaxy}.
\newblock \emph{Astrophys. J. Lett.}, 635:\penalty0 L37--L40, 2005.
\newblock \doi{10.1086/499161}.

\bibitem[McConnachie and Irwin(2006)]{McConnachie:2005td}
Alan McConnachie and Mike Irwin.
\newblock {Structural parameters for the m31 dwarf spheroidals}.
\newblock \emph{Mon. Not. Roy. Astron. Soc.}, 365:\penalty0 1263, 2006.
\newblock \doi{10.1111/j.1365-2966.2005.09806.x}.

\bibitem[{van der Marel} et~al.(2019){van der Marel}, {Fardal}, {Sohn},
  {Patel}, {Besla}, {del Pino}, {Sahlmann}, and {Watkins}]{vdm2019}
Roeland~P. {van der Marel}, Mark~A. {Fardal}, Sangmo~Tony {Sohn}, Ekta {Patel},
  Gurtina {Besla}, Andr{\'e}s {del Pino}, Johannes {Sahlmann}, and Laura~L.
  {Watkins}.
\newblock {First Gaia Dynamics of the Andromeda System: DR2 Proper Motions,
  Orbits, and Rotation of M31 and M33}.
\newblock \emph{\apj}, 872\penalty0 (1):\penalty0 24, February 2019.
\newblock \doi{10.3847/1538-4357/ab001b}.

\bibitem[{van der Marel} et~al.(2012){van der Marel}, {Besla}, {Cox}, {Sohn},
  and {Anderson}]{vdm2012}
Roeland~P. {van der Marel}, Gurtina {Besla}, T.~J. {Cox}, Sangmo~Tony {Sohn},
  and Jay {Anderson}.
\newblock {The M31 Velocity Vector. III. Future Milky Way M31-M33 Orbital
  Evolution, Merging, and Fate of the Sun}.
\newblock \emph{\apj}, 753\penalty0 (1):\penalty0 9, July 2012.
\newblock \doi{10.1088/0004-637X/753/1/9}.

\bibitem[van~der Marel and Guhathakurta(2008)]{vanderMarel:2007yw}
Roeland~P. van~der Marel and Puragra Guhathakurta.
\newblock {M31 Transverse Velocity and Local Group Mass from Satellite
  Kinematics}.
\newblock \emph{Astrophys. J.}, 678:\penalty0 187, 2008.
\newblock \doi{10.1086/533430}.

\bibitem[Skilling et~al.(2006)]{skilling2006nested}
John Skilling et~al.
\newblock Nested sampling for general bayesian computation.
\newblock \emph{Bayesian analysis}, 1\penalty0 (4):\penalty0 833--859, 2006.

\bibitem[{Handley} et~al.(2015{\natexlab{a}}){Handley}, {Hobson}, and
  {Lasenby}]{Handley:2015a}
W.~J. {Handley}, M.~P. {Hobson}, and A.~N. {Lasenby}.
\newblock {POLYCHORD: nested sampling for cosmology}.
\newblock \emph{\mnras}, 450:\penalty0 L61--L65, June 2015{\natexlab{a}}.
\newblock \doi{10.1093/mnrasl/slv047}.

\bibitem[{Handley} et~al.(2015{\natexlab{b}}){Handley}, {Hobson}, and
  {Lasenby}]{Handley:2015b}
W.~J. {Handley}, M.~P. {Hobson}, and A.~N. {Lasenby}.
\newblock {POLYCHORD: next-generation nested sampling}.
\newblock \emph{\mnras}, 453:\penalty0 4384--4398, November 2015{\natexlab{b}}.
\newblock \doi{10.1093/mnras/stv1911}.

\bibitem[Handley(2019)]{anesthetic}
Will Handley.
\newblock anesthetic: nested sampling visualisation.
\newblock \emph{The Journal of Open Source Software}, 4\penalty0 (37):\penalty0
  1414, Jun 2019.
\newblock \doi{10.21105/joss.01414}.

\bibitem[O'Brien et~al.(2016)O'Brien, Kashinath, Cavanaugh, Collins, and
  O'Brien]{fastkde1}
Travis O'Brien, Karthik Kashinath, Nicholas Cavanaugh, William Collins, and
  John O'Brien.
\newblock A fast and objective multidimensional kernel density estimation
  method: Fastkde.
\newblock \emph{Computational Statistics \& Data Analysis}, 101, 03 2016.
\newblock \doi{10.1016/j.csda.2016.02.014}.

\bibitem[O’Brien et~al.(2014)O’Brien, Collins, Rauscher, and
  Ringler]{fastkde2}
Travis O’Brien, William Collins, Sara Rauscher, and Todd Ringler.
\newblock Reducing the computational cost of the ecf using a nufft: A fast and
  objective probability density estimation method.
\newblock \emph{Computational Statistics \& Data Analysis}, 79, 11 2014.
\newblock \doi{10.1016/j.csda.2014.06.002}.

\bibitem[{Press} and {Schechter}(1974)]{Press1974}
William~H. {Press} and Paul {Schechter}.
\newblock {Formation of Galaxies and Clusters of Galaxies by Self-Similar
  Gravitational Condensation}.
\newblock \emph{\apj}, 187:\penalty0 425--438, February 1974.
\newblock \doi{10.1086/152650}.

\bibitem[Diemer(2018)]{Diemer:2017bwl}
Benedikt Diemer.
\newblock {COLOSSUS: A python toolkit for cosmology, large-scale structure, and
  dark matter halos}.
\newblock \emph{Astrophys. J. Suppl.}, 239\penalty0 (2):\penalty0 35, 2018.
\newblock \doi{10.3847/1538-4365/aaee8c}.

\bibitem[Tinker et~al.(2008)Tinker, Kravtsov, Klypin, Abazajian, Warren, Yepes,
  Gottlober, and Holz]{Tinker:2008ff}
Jeremy~L. Tinker, Andrey~V. Kravtsov, Anatoly Klypin, Kevork Abazajian,
  Michael~S. Warren, Gustavo Yepes, Stefan Gottlober, and Daniel~E. Holz.
\newblock {Toward a halo mass function for precision cosmology: The Limits of
  universality}.
\newblock \emph{Astrophys. J.}, 688:\penalty0 709--728, 2008.
\newblock \doi{10.1086/591439}.

\bibitem[Skilling()]{skilling2006}
John Skilling.
\newblock Nested sampling for general bayesian computation.
\newblock \emph{Bayesian Anal.}, \penalty0 (4):\penalty0 833--859, 12 .
\newblock \doi{10.1214/06-BA127}.

\bibitem[{Planck Collaboration}(2020)]{planck_2018_VI}
{Planck Collaboration}.
\newblock Planck 2018 results - vi. cosmological parameters.
\newblock \emph{A\&A}, 641:\penalty0 A6, 2020.
\newblock \doi{10.1051/0004-6361/201833910}.

\bibitem[{Libeskind} et~al.(2020){Libeskind}, {Carlesi}, {Grand}, {Khalatyan},
  {Knebe}, {Pakmor}, {Pilipenko}, {Pawlowski}, {Sparre}, {Tempel}, {Wang},
  {Courtois}, {Gottl{\"o}ber}, {Hoffman}, {Minchev}, {Pfrommer}, {Sorce},
  {Springel}, {Steinmetz}, {Tully}, {Vogelsberger}, and
  {Yepes}]{2020MNRAS.498.2968L}
Noam~I. {Libeskind}, Edoardo {Carlesi}, Robert J.~J. {Grand}, Arman
  {Khalatyan}, Alexander {Knebe}, Ruediger {Pakmor}, Sergey {Pilipenko},
  Marcel~S. {Pawlowski}, Martin {Sparre}, Elmo {Tempel}, Peng {Wang},
  H{\'e}l{\`e}ne~M. {Courtois}, Stefan {Gottl{\"o}ber}, Yehuda {Hoffman}, Ivan
  {Minchev}, Christoph {Pfrommer}, Jenny~G. {Sorce}, Volker {Springel},
  Matthias {Steinmetz}, R.~Brent {Tully}, Mark {Vogelsberger}, and Gustavo
  {Yepes}.
\newblock {The HESTIA project: simulations of the Local Group}.
\newblock \emph{\mnras}, 498\penalty0 (2):\penalty0 2968--2983, August 2020.
\newblock \doi{10.1093/mnras/staa2541}.

\bibitem[{Diaz} et~al.(2014){Diaz}, {Koposov}, {Irwin}, {Belokurov}, and
  {Evans}]{Diaz2014}
J.~D. {Diaz}, S.~E. {Koposov}, M.~{Irwin}, V.~{Belokurov}, and N.~W. {Evans}.
\newblock {Balancing mass and momentum in the Local Group}.
\newblock \emph{\mnras}, 443\penalty0 (2):\penalty0 1688--1703, September 2014.
\newblock \doi{10.1093/mnras/stu1210}.

\bibitem[{Zaritsky} and {Courtois}(2017)]{Zaritsky2017}
Dennis {Zaritsky} and Helene {Courtois}.
\newblock {A dynamics-free lower bound on the mass of our Galaxy}.
\newblock \emph{\mnras}, 465\penalty0 (3):\penalty0 3724--3728, March 2017.
\newblock \doi{10.1093/mnras/stw2922}.

\bibitem[Hattori et~al.(2018)Hattori, Valluri, Bell, and
  Roederer]{Hattori:2018isv}
Kohei Hattori, Monica Valluri, Eric~F. Bell, and Ian~U. Roederer.
\newblock {Old, Metal-Poor Extreme Velocity Stars in the Solar Neighborhood}.
\newblock \emph{Astrophys. J.}, 866\penalty0 (2):\penalty0 121, 2018.
\newblock \doi{10.3847/1538-4357/aadee5}.

\bibitem[{Posti} and {Helmi}(2019)]{Posti2019}
Lorenzo {Posti} and Amina {Helmi}.
\newblock {Mass and shape of the Milky Way's dark matter halo with globular
  clusters from Gaia and Hubble}.
\newblock \emph{\aap}, 621:\penalty0 A56, January 2019.
\newblock \doi{10.1051/0004-6361/201833355}.

\bibitem[{Watkins} et~al.(2019){Watkins}, {van der Marel}, {Sohn}, and
  {Evans}]{Watkins2019}
Laura~L. {Watkins}, Roeland~P. {van der Marel}, Sangmo~Tony {Sohn}, and N.~Wyn
  {Evans}.
\newblock {Evidence for an Intermediate-mass Milky Way from Gaia DR2 Halo
  Globular Cluster Motions}.
\newblock \emph{\apj}, 873\penalty0 (2):\penalty0 118, March 2019.
\newblock \doi{10.3847/1538-4357/ab089f}.

\bibitem[Karukes et~al.(2020)Karukes, Benito, Iocco, Trotta, and
  Geringer-Sameth]{Karukes:2019jwa}
Ekaterina~V. Karukes, Maria Benito, Fabio Iocco, Roberto Trotta, and Alex
  Geringer-Sameth.
\newblock {A robust estimate of the Milky Way mass from rotation curve data}.
\newblock \emph{JCAP}, 05:\penalty0 033, 2020.
\newblock \doi{10.1088/1475-7516/2020/05/033}.

\bibitem[{Corbelli} et~al.(2010){Corbelli}, {Lorenzoni}, {Walterbos}, {Braun},
  and {Thilker}]{Corbelli2010}
E.~{Corbelli}, S.~{Lorenzoni}, R.~{Walterbos}, R.~{Braun}, and D.~{Thilker}.
\newblock {A wide-field H I mosaic of Messier 31. II. The disk warp, rotation,
  and the dark matter halo}.
\newblock \emph{\aap}, 511:\penalty0 A89, February 2010.
\newblock \doi{10.1051/0004-6361/200913297}.

\bibitem[{Tamm} et~al.(2012){Tamm}, {Tempel}, {Tenjes}, {Tihhonova}, and
  {Tuvikene}]{Tamm2012}
A.~{Tamm}, E.~{Tempel}, P.~{Tenjes}, O.~{Tihhonova}, and T.~{Tuvikene}.
\newblock {Stellar mass map and dark matter distribution in M 31}.
\newblock \emph{\aap}, 546:\penalty0 A4, October 2012.
\newblock \doi{10.1051/0004-6361/201220065}.

\bibitem[Kafle et~al.(2018)Kafle, Sharma, Lewis, Robotham, and
  Driver]{Kafle:2018amm}
Prajwal~R. Kafle, Sanjib Sharma, Geraint~F. Lewis, Aaron S.~G. Robotham, and
  Simon~P. Driver.
\newblock {The Need for Speed: Escape velocity and dynamical mass measurements
  of the Andromeda galaxy}.
\newblock \emph{Mon. Not. Roy. Astron. Soc.}, 475\penalty0 (3):\penalty0
  4043--4054, 2018.
\newblock \doi{10.1093/mnras/sty082}.

\bibitem[{Pe{\~n}arrubia} et~al.(2016){Pe{\~n}arrubia}, {G{\'o}mez}, {Besla},
  {Erkal}, and {Ma}]{2016MNRAS.456L..54P}
Jorge {Pe{\~n}arrubia}, Facundo~A. {G{\'o}mez}, Gurtina {Besla}, Denis {Erkal},
  and Yin-Zhe {Ma}.
\newblock {A timing constraint on the (total) mass of the Large Magellanic
  Cloud}.
\newblock \emph{\mnras}, 456\penalty0 (1):\penalty0 L54--L58, February 2016.
\newblock \doi{10.1093/mnrasl/slv160}.

\bibitem[Theis and Bethge(2015)]{theis2015generative}
Lucas Theis and Matthias Bethge.
\newblock Generative image modeling using spatial lstms.
\newblock In \emph{Advances in Neural Information Processing Systems}, pages
  1927--1935, 2015.

\bibitem[{Salimans} et~al.(2017){Salimans}, {Karpathy}, {Chen}, and
  {Kingma}]{Salimans2017}
Tim {Salimans}, Andrej {Karpathy}, Xi~{Chen}, and Diederik~P. {Kingma}.
\newblock {PixelCNN++: Improving the PixelCNN with Discretized Logistic Mixture
  Likelihood and Other Modifications}.
\newblock \emph{arXiv e-prints}, art. arXiv:1701.05517, January 2017.

\bibitem[Lewis(2019)]{Lewis:2019xzd}
Antony Lewis.
\newblock {GetDist: a Python package for analysing Monte Carlo samples}.
\newblock 2019.
\newblock URL \url{https://getdist.readthedocs.io}.

\bibitem[{Hinton}(2016)]{Hinton2016}
S.~R. {Hinton}.
\newblock {ChainConsumer}.
\newblock \emph{The Journal of Open Source Software}, 1:\penalty0 00045, August
  2016.
\newblock \doi{10.21105/joss.00045}.

\bibitem[Dempster et~al.(1977)Dempster, Laird, and Rubin]{Dempster1977}
A.~P. Dempster, N.~M. Laird, and D.~B. Rubin.
\newblock Maximum likelihood from incomplete data via the em algorithm.
\newblock \emph{Journal of the Royal Statistical Society. Series B
  (Methodological)}, 39\penalty0 (1):\penalty0 1--38, 1977.
\newblock ISSN 00359246.
\newblock URL \url{http://www.jstor.org/stable/2984875}.

\bibitem[Germain et~al.(2015)Germain, Gregor, Murray, and
  Larochelle]{germain2015made}
Mathieu Germain, Karol Gregor, Iain Murray, and Hugo Larochelle.
\newblock Made: Masked autoencoder for distribution estimation.
\newblock In \emph{International Conference on Machine Learning}, pages
  881--889, 2015.

\bibitem[Rayleigh(1905)]{rayleigh1905problem}
Lord Rayleigh.
\newblock The problem of the random walk.
\newblock \emph{Nature}, 72\penalty0 (1866):\penalty0 318, 1905.

\end{thebibliography}

\end{document}